\documentclass{aa}  
\usepackage{graphicx}
\usepackage{longtable,lscape}
\usepackage[varg]{txfonts}
\usepackage{natbib}

\newcommand{\ks}{$K_s$}
\newcommand{\rmz}{$R-z'$}
\newcommand{\zmk}{$z'-K_s$}
\newcommand{\rzk}{$Rz'K_s$ }
\newcommand{\micron}{\,$\mu m$}

\begin{document}

   \title{A multi-wavelength survey of AGN in the XMM-LSS field:\\
   I. Quasar selection via the $KX$ technique}

   \author{Th. Nakos         \inst{1,2}
      \and J.P. Willis       \inst{3}
      \and S. Andreon        \inst{4}
      \and J. Surdej         \inst{2} \fnmsep\thanks{also Directeur de Recherches Honoraire FNRS}
      \and P. Riaud          \inst{2,5}  
      \and E. Hatziminaoglou \inst{6} 
      \and O. Garcet         \inst{2}
      \and D. Alloin         \inst{7}
      \and M. Baes           \inst{1}
      \and G. Galaz          \inst{8}
      \and M. Pierre         \inst{7}
      \and H. Quintana       \inst{8}
      \and M.J. Page         \inst{9}
      \and J.A. Tedds        \inst{10}
      \and M.T. Ceballos     \inst{11}
      \and A. Corral         \inst{11}
      \and J. Ebrero         \inst{11}
      \and M. Krumpe         \inst{12}
      \and S. Mateos         \inst{10}
                                  }
   \offprints{Th. Nakos (theodoros.nakos@ugent.be)}

   \institute{Sterrenkundig Observatorium, University of Ghent, Krijgslaan 281 S9, B-9000  Ghent, Belgium
     \and
          Institut d'Astrophysique et de G\'eophysique, Universit\'e de Li\`ege, 
	      All\'ee du 6 Ao{\^u}t, 17, B5c, B-4000 Sart Tilman, Belgium
	 \and   
	      Department of Physics and Astronomy, Univ. of Victoria, Elliott Building 3,
	      800 Finnerty Road, Victoria, BC, V8P 1A1 Canada 
	 \and
	      Osservatorio Astronomico di Brera, via Brera 28, 20121 Milano, Italy
     \and
	      60 rue des Bergers, 75015 Paris, France 
	 \and
	      European Southern Observatory, Karl-Schwarzschild-Str. 2, D-85748 Garching bei M\"{u}nchen, Germany
     \and
          Laboratoire AIM, CEA/DSM-CNRS-Universit\'e Paris Diderot, 
          IRFU/Service d'Astrophysique, B{\^a}t. 709, CEA Saclay, 91191, Gif sur Yvette C\'edex, France
     \and
	     Departamento de Astronom{\'i}a y Astrof{\'i}sica, Pontificia Universidad Cat{\'o}lica de Chile, 
	     Casilla 306, Santiago 22, Chile
     \and 
         Mullard Space Science Laboratory, University College London, Holmbury St Mary, Dorking, Surrey, RH5 6NT, UK
     \and 
         Department of Physics and Astronomy, University of Leicester, LE1 7RH, UK     
     \and
         Instituto de F\'\i sica de Cantabria (CSIC-UC), 39005 Santander, Spain 
     \and
         Astrophysikalisches Institut Potsdam, An der Sternwarte 16, 14482 Potsdam, Germany
                }

   \date{Received ; accepted }

 
  \abstract
   {}
   {We present a sample of candidate quasars selected using the
     $KX$-technique. The data cover $0.68 \deg^2$ of the X-ray
     Multi-Mirror (XMM) Large-Scale Structure (LSS) survey area where
     overlapping multi-wavelength imaging data permits an
     investigation of the physical nature of selected sources.}
   {The $KX$ method identifies quasars on the basis of their optical
     ($R$ and $z'$) to near-infrared (\ks) photometry and point-like
     morphology. We combine these data with optical ($u^*,g'r',i',z'$)
     and mid-infrared ($3.6-24\, \mu m$) wavebands to reconstruct the
     spectral energy distributions (SEDs) of candidate quasars.}
   {Of 93 sources selected as candidate quasars by the $KX$ method, 25
     are classified as quasars by the subsequent SED
     analysis. Spectroscopic observations are available for 12/25 of
     these sources and confirm the quasar hypothesis in each
     case.  Even more, 90\% of the SED-classified quasars show X-ray emission,
     a property not shared by any of the false candidates in the KX-selected
     sample. Applying a photometric redshift analysis to the sources
     without spectroscopy indicates that the 25 sources classified as
     quasars occupy the interval $0.7 \le z \le 2.5$. The remaining
     68/93 sources are classified as stars and unresolved galaxies.  }
   {}

   \titlerunning{A multi-wavelength survey for AGN in the XMM-LSS field: I. Quasar selection via the $KX$ technique}
   \maketitle

%
\section{Introduction}

Despite the enormous progress that wide-field imaging surveys have
provided in understanding quasar populations over the past decade,
concerns remain over the extent to which observational biases
contribute to a partial view of the quasar phenomenon.  Consequently,
multi-wavelength observations are important not only to reproduce the
spectral energy distribution (SED) of the objects under study, but
also to provide as complete a census as is practical of the quasar
population.  Nevertheless, selecting source populations using
different energy bands encompasses a risk of confusion between
different types of objects or different physical processes. Thus,
studying the selection effects is an essential step to
understanding the properties of the parent population.

A classic method to identify type-1 quasars in optical imaging surveys
is known as the $UV$-excess ($UVX$; \citealt{sandage65},
\citealt{schmidt83}). The $UVX$ technique
exploits the fact that quasars display an emission excess in blue
wavebands compared to main sequence, late type Galactic stars,
and therefore occupy a distinct (i.e. bluer) locus in a color-color
diagram with respect to stars.

Based on a study of 323 radio-loud sources selected from the Parkes
catalog~\citep{parks90}, approximately 70\% of which were identified
spectroscopically as quasars, \cite{webster95} claimed that up to 80\%
of quasars could be missed in optical surveys. Their suggestion was
based on the $B_J - K$ color index of these radio-selected sources,
which showed a much larger scatter ($1 < B_J - K < 8$) compared to the
typical values for optically selected quasars.  Various
interpretations were suggested, such as (a) intrinsic light
absorption, (b) absorption from dust along the line of sight or (c)
differences in the internal processes in the active nucleus.  Cases
(a) and (b) would result in ``reddened" quasars, whilst objects in
category (c) would be considered as intrinsically ``red" (in contrast
to the ``classical", blue type-1 quasars).

However, a number of studies have questioned the dusty quasar
hypothesis.  \cite{masci98} rejected the dust hypothesis
and~\cite{whiting01} suggesting that the $B_J-K$ spread could be
accounted for by synchrotron emission. \cite{benn98} claimed that the
$B_J - K$ scatter could be attributed to observational effects such as
variability, underestimated photometric uncertainties, etc.
Although~\cite{francis00} support the conclusions
of~\cite{webster95}, \cite{richards03} demonstrate that, for the
majority of the SDSS quasars, redder colors could be explained by
intrinsic processes in the AGN and that the number of missed red and
reddened quasars in surveys such as the SDSS (different from typical
optical surveys because of the filter selection) comes down to
``only'' 15\%. \cite{brown06} found similar results, suggesting that
red \mbox{type-1} quasars correspond to about 20\% of the type-1
population.

With the advent of large field, near infrared (NIR) surveys,
\cite{warren00} suggested a new colour excess criterion, designed to
be less prone to the problem of dust reddening and extinction. Their
method is the NIR analog of $UVX$: as the quasar SED follows a
power-law, it displays a flux excess at NIR wavelengths compared to
the Rayleigh-Jeans tail observed in stars. As a result quasars should
also occupy a distinct region of the NIR color-color diagram, as they
do in the $UBV$ color-color plane. The so-called $K$-excess criterion,
hereafter $KX$, would serve as an alternative to the $UVX$ for
selecting quasars at redshifts $z > 2.2$. At these redshifts, the
UV/optical colors of quasars become virtually indistinguishable from
those of late type stars.

Despite being a prominsing technique, relatively few $KX$-selected
quasar samples have appeared in the literature \---\ mainly due to the
considerable effort required to complete a large area NIR
imaging survey. The conclusions drawn from the $KX$-selected quasar
populations published to date are rather limited as various
factors such as the small area of the survey~\citep{croom01}, the
relative shallow magnitudes sampled~\citep{barkhouse01} or the
inhomogeneity and incompleteness of the survey~\citep{sharp02} have
hampered an assessment of the potential of the method. The first
detailed study of the $KX$ method was given by~\cite{jurek07}
while more recently~\cite{smail08} also published a work on
$KX$-selected quasars, raising the problem of contamination by 
foreground compact galaxies and their impact on spectroscopic
follow-up observations.

In this paper we present an analysis of a sample of 93 $KX$-selected
quasar candidates selected from 0.68 deg.$^2$ of \rzk\ imaging data
located in the X-ray Multi-Mirror (XMM) Large Scale Structure (LSS)
survey field \citep{pierre04}.  The XMM LSS survey is an X-ray
imaging survey covering approximately $10 \deg^2$ to an approximate
point-source flux limit of $8\times 10^{-15} \rm erg s^{-1} cm^{-2}$
in the [0.5-2]~keV energy band.  The XMM LSS is associated with a
number of imaging data sets at different wavebands: the XMM LSS
contains the XMM Medium Deep Survey (XMDS), a deeper X-ray imaging
component over 2 deg.$^2$ in addition to the optical
Canada-France-Hawaii Telescope Legacy Survey (CFHTLS) and the Spitzer
Wide area IR Extragalactic (SWIRE) survey.  These additional data
provide a panchromatic view of the quasar candidates and permit a
detailed discussion of their physical nature.

The paper is organized as follows: Section 2 presents the optical and
NIR data used in the $KX$ technique. Section 3 describes the
multi-wavelength data available in the field. Section 4 describes the
$KX$ selection criteria and discusses the physical nature of the
selected sources using their multi-wavelength properties.  In Section
5 we draw some conclusions on the color properties of the confirmed
quasars selected using $KX$ and comment on the overall efficiency of
the technique. We also consider some of the challenges to be overcome
in applying the $KX$ technique to large data sets.

%
\section{Optical and near-infrared observations and data reduction}
%
Two data sets were combined to produce a single optical plus
near-infrared catalog upon which to conduct a search for quasars
using the $KX$ technique. Optical $R,z'$ data were obtained at the
Cerro Tololo Interamerican Observatory (CTIO) and near-infrared
\ks-band data were obtained at Las Campanas Observatory (LCO).  The
two data sets were reduced independently and subsequently matched to
form a single catalog. In this section we briefly describe the
observations, data reduction and general properties of these
photometric data sets.

%
\subsection{Near-infrared observations}
%
Near-infrared observations were performed during the period
25\,$-$\,28 October 2002 using the 2.5\,m (100 inch) Du Pont
telescope, at Las Campanas Observatory.  The XMDS field was observed
in the \ks-band, using the Wide Field Infrared Camera
(WIRC,~\citealt{persson02}).  WIRC is a NIR ($1.0-2.5\,\mu$m) camera
consisting of four $1024^2$ Rockwell HAWAII arrays. Each array covers
an area $3.4 \times 3.4$ arcmin$^2$ with a scale of 0.2$''$/pixel. The
four detectors are equally spaced with respect to the center of the
configuration, with a gap between them of about 3.05$\arcmin$.

Each telescope pointing consisted of a regular grid of nine dithering
pointings with a 20$\arcsec$ offset.  A total of $4 \times 28 \rm s$
exposures were obtained at each dithered location and each combined
image displays a total integration time of $ 9 \times 4 \times 28 \rm
s = 1008s$ in the central area.  The nights of Oct. 25 and Oct. 27
were of mediocre photometric quality, while the nights of Oct. 26 and Oct. 28 were
photometric.  The typical seeing conditions during
the run were between $1.1-1.5''$ FWHM.  Several hundred images
were obtained and then combined to generate a mosaic of 0.8
$\deg^2$.

\subsection{Optical observations}
The full XMDS region was imaged in the $R$- and $z'$-bands using the
Mosaic II CCD imager mounted on the 4\,m Blanco telescope at CTIO
observatory. The Mosaic II camera consists of eight $2048 \times 4096$
CCD arrays. The scale at the center of the camera is 0.27$''$/pixel,
covering in total $36' \times 36'$ on the sky.
The observations were obtained in November 2001 during photometric
conditions with an average seeing of FWHM\,$\approx 1.3\arcsec$.
Standard procedures were followed for the reduction of the obtained
images.

The CTIO individual frames were astrometrically calibrated using the
USNO-2 catalog~\citep{monet03} in order to generate mosaic images of
the observed area. The final astrometric precision reached was on
the order of 0.3$''$. Source extraction and photometry were performed
using {\tt SExtractor}~\citep{bertin96} with independent catalogs
generated for each band.  Detailed information regarding the data
processing of the CTIO data can be found in~\cite{andreon04}.

%
\subsection{\label{drp}Near-infrared data processing}
%
Near-infrared data were processed using a set of scripts based upon
the {\sf IRAF}\footnote{IRAF is distributed by the National Optical
  Astronomy Observatory, which is operated by the Association of
  Universities for Research in Astronomy, Inc., under a cooperative
  agreement with the National Science Foundation.}  and {\sf
  PHIIRS}~\citep{hall98} packages. The reduction strategy is based
upon previous near-infrared surveys reported by ~\cite{chen02} and
\cite{labbe03}. A comprehensive description of the pipeline is given by
\cite{nakos07} and in this paper we provide a brief description of the
data processing sequence.
\begin{enumerate}
\item The sky background, residual features and electronic signatures
  were removed from individual (dithered) frames employing a
  combination of comparison frames from which objects had been masked.
\item Individual frames were corrected to a uniform photometric
  sensitivity. Corrections are required to (a) bring all four WIRC
  detectors to the same sensitivity level, (b) remove differential
  airmass effects, and (c) calibrate all frames to a standard
  photometric system. For this purpose, six photometric standard
  stars~\citep{persson98} were observed during the night of October
  26.
  For the two photometric nights these observations were employed to
  calculate a zero point and extinction coefficient.  For the
  non-photometric nights, the 2MASS photometry~\citep{skrutskie06} was
  used to set the zero point that would shift the instrumental
  \ks-band magnitudes to calibrated \ks-band photometry. Full details
  about the calibration can be found in~\cite{nakos07}.
\item An astrometric solution was determined for the dithered frames
  and applied to generate a mosaic image. This was achieved by
  comparing the coordinates of point-like sources present in the \ks,
  2MASS and $z'$-band frames.  Having determined the astrometric
  solution, individual frames were combined to generate four mosaic
  images.
\item The source completeness and false detection rate in the mosiac
  images were computed using simulations to determine a suitable
  magnitude limit in the \ks-band catalog. As the \ks\, catalog is 
  later matched to the $R$ and $z'$ photometric catalogs it is the
  \ks-band 's faintest magnitude that determines the \rzk survey's
  sensitivity limit.
\end{enumerate}

\begin{figure}
\centering
\includegraphics[width=9cm]{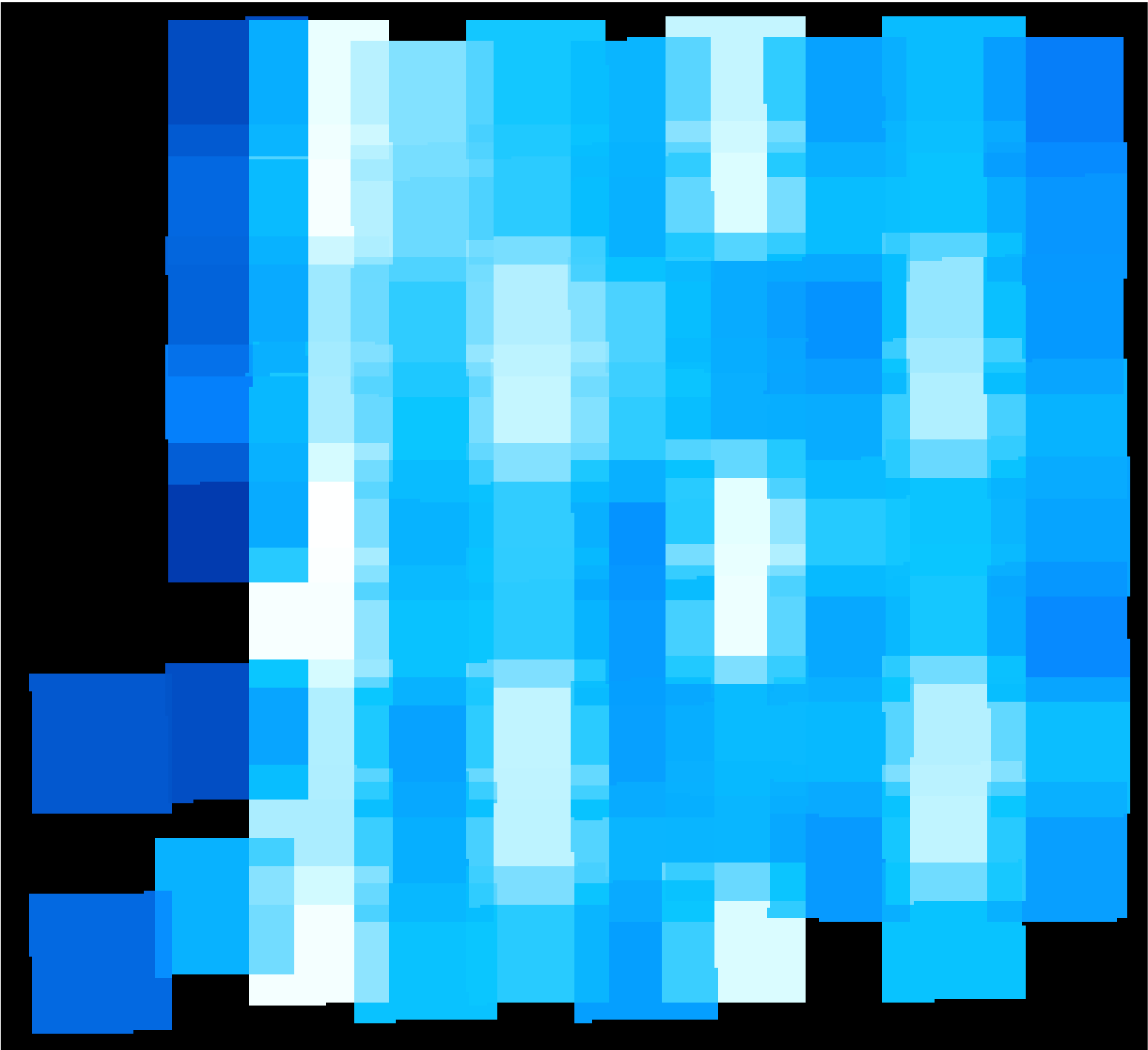}
\caption{Composite exposure map, generated by combining about 75 individual 
exposure maps in which the photometric calibration effects 
(zero point, airmass) have also been incorporated. By dividing the science
mosaic by its corresponding composite exposure map we 
directly derive the calibrated magnitude of the sources in the field. The ``sensitivity''
of the lighter areas is three times higher to that of the dark
areas. Sources lying in the
overlapping regions of the mosaic (formed by adjacent pointings)  have 
a better signal-to-noise ratio.}
\label{mosaic-expo}
\end{figure}

Analysis of the mosaiced images indicated that the impact of the
reduction procedure on the photometric uncertainty was at the level of
a few hundredths of a magnitude.  Exposure mosaics indicating the
exact exposure time of pixels in the associated science mosaic were
also produced.  These frames were used to convert the instrumental
magnitudes measured on the science frames into magnitudes in the Vega
photometric system. An exposure map of a sub-region of the LCO mosaic
image is presented in Fig.~\ref{mosaic-expo}.  The sensitivity ratio
between the deepest exposure maps (white) and the shallowest
(dark-colored) is on the order of three.

\begin{figure*}
\centering
\includegraphics[width=13cm]{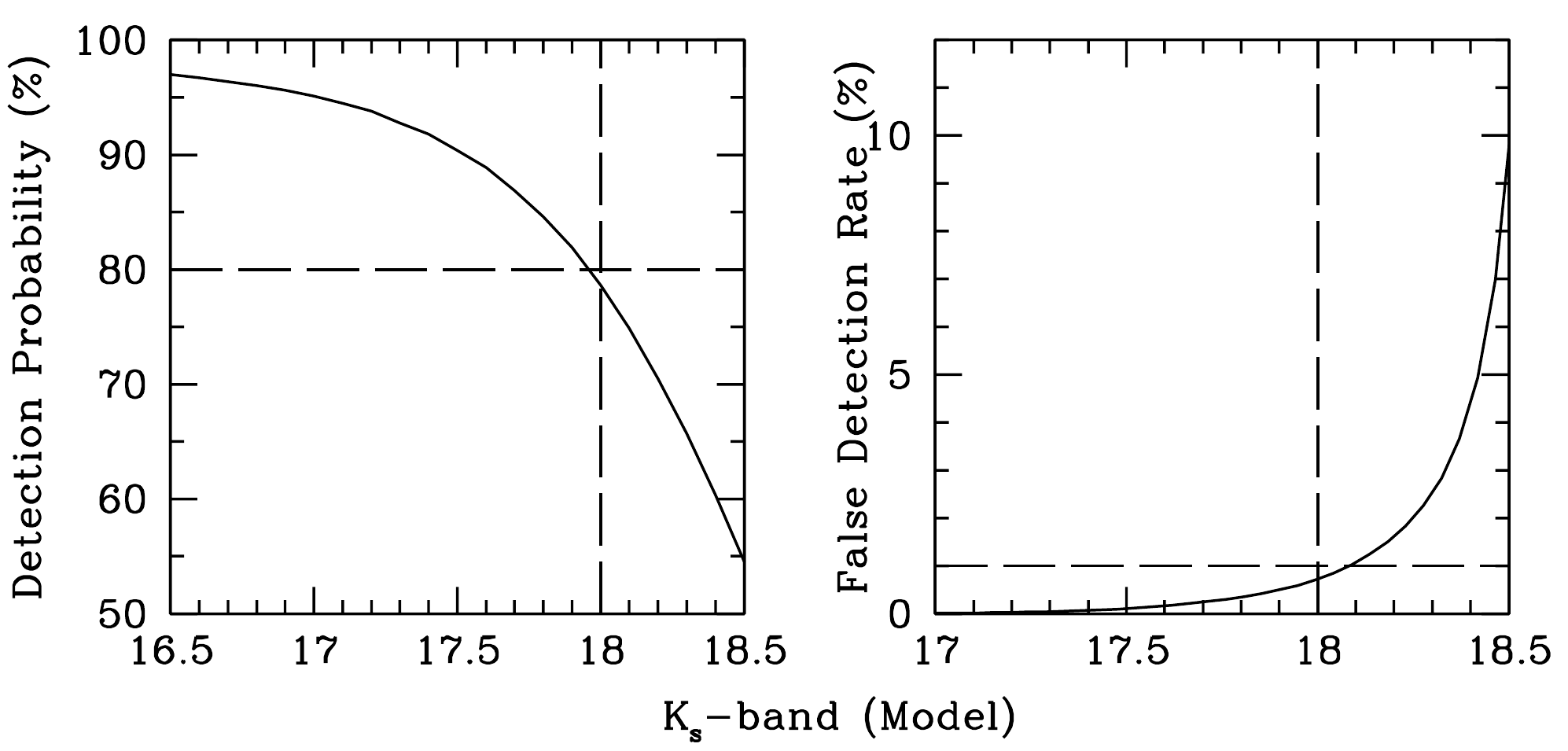}
\caption{
{\it Left panel:} Average detection probability, over the four science mosaics, 
as a function of the \ks\,magnitude (see text for more details). 
{\it Right panel:} Average false detection rate as a function of the \ks\,magnitude.
The dashed lines indicate the completeness (left panel) and expected percentage
of false detections (right panel) for a catalog cut at $K_s =18$.  
}
\label{simul}
\end{figure*}

\subsection{\label{simul-sex}{Constructing the \ks-band catalog}}

Figure \ref{mosaic-expo} indicates that the total exposure time, and
therefore the background noise, is not constant over the extent of a
given mosaic.  Therefore understanding the spatial noise properties
of the mosaic is essential for generating the \ks-band object catalog.
The limiting source brightness is determined considering: (a) the
probability of detecting a real source, (b) the number of false
detections as a function of source brightness. The first condition is
required to determine the percentage of sources which are missed as a
function of the limiting magnitude of the survey, while the second
condition determines the balance between a faint brightness threshold,
which will introduce many spurious detections, and a bright threshold
which will exclude a significant number of interesting objects.

This study was performed using a code, kindly provided to us by
E. Labbe (PUC) and F.  Courbin (Ecole Polytechnique F\'ed\'erale de
Lausanne, Switzerland), that simulates the presence of unresolved
sources in our data.  Artificial sources are introduced into each
science mosaic in a regular grid of locations modulo a small random
offset that varies with each simulation. The typical sky area
associated with each grid location is 10$\arcsec$.  Source extraction
is performed on each simulated image and the relative frequency with
which the artificial sources are recovered is stored as the detection
probability. This process is then repeated for successively fainter
artificial sources. The results of the simulations, run on all four
mosaics, were averaged in order to obtain a mean value for the
detection probability and for the false detection rate. Based on these
averaged results, presented in Fig.~\ref{simul}, we concluded that
excluding detections fainter than $K_s = 18$ (within a Kron-type aperture) 
from the final \ks-band catalog provides a good balance between a
low percentage of false detections (on the order of a few percent) and
a satisfactory completeness of the \ks\, catalog (80\%).  In addition to
a magnitude cut, sources with erroneous photometry were removed from
the catalog using a combination of {\tt SExtractor} flag information
and visual inspection. Such rejected sources were typically: (a)
sources suffering from blending effects due to the presence of a
counterpart within a few arcseconds, (b) sources close to saturated
objects, (c) sources close to the mosaic borders, or (d) sources
affected by bad columns or other artifacts.

The final \ks-band catalog contains the position (RA,Dec) and the
fixed aperture (3$''$ diameter) photometry of some $\sim$4500 entries.
These entries were selected according to: (a) their Kron magnitude (the
one used during the simulations), and (b) the detection probability
corresponding to the position of each detection in the mosaic images
(as derived from the simulations).

\subsection{\label{rzk} Matching the \rzk catalogs}

Source extraction and photometry was performed using {\tt SExtractor}
for the optical $R$ and $z'$ image mosaics using similar parameters
and checks as used for the \ks\ data. Source photometry in each band
was combined into a single catalog by matching the source astrometry
in each band within a specified $2\arcsec$ tolerance.  The tolerance
value was specified considering the astrometric precision of the
catalogs and the object density in the LCO field: $R,z'$ counterparts
were identified for all \ks-band detections, even those with a poorer
astrometric accuracy than average (0.5\arcsec), with a minimum risk of
misidentification among the few cases of blended objects that were not
removed throughout the ``cleaning'' process described above.  The
final matched catalog contains some 3400 sources and general catalog
properties are presented in Table~\ref{general}.

Matched sources were classified as point-like or extended on the basis
of the $R$- and $z'$-band magnitude and the {\tt SExtractor}
stellarity index.  It should be noted that although the stellarity
index provides reliable information concerning the morphology of
relatively bright sources (typical magnitudes $R,z' < 22$), it is less
relaible when considering fainter sources. A combination of the
stellarity index and source magnitude provides a more reliable
morphological classification. Additional details can be found
in~\cite{andreon04}.  Out of the 3400 sources in the matched \rzk
catalog some 2450 ($\sim 72$\%) were classified as extended, while
the remainder were classified as point-like.

\begin{table}
\caption{General properties of the \ks-band selected \rzk survey. 
The astrometric and photometric precision are quoted at $1\,\sigma$.}
\label{general}
\centering
\begin{tabular}{|c|c|c|c|}
\hline
  {\bf Filter} &  {\bf Astr. prec.} & {\bf Phot. prec.} & {\bf Mag limit}\\
   \hline
   \hline
   $R$         &     0.3$''$  	 &     0.02	  &	     22.5\\
   $z'$        &     0.3$''$  	 &     0.02	  &	     21.0\\
   $K_s$       &     0.5$''$   	 &     0.07	  &	     18.0\\
   \hline
\end{tabular}
\end{table}

\begin{figure}[h]
\centering
\includegraphics[width=9cm]{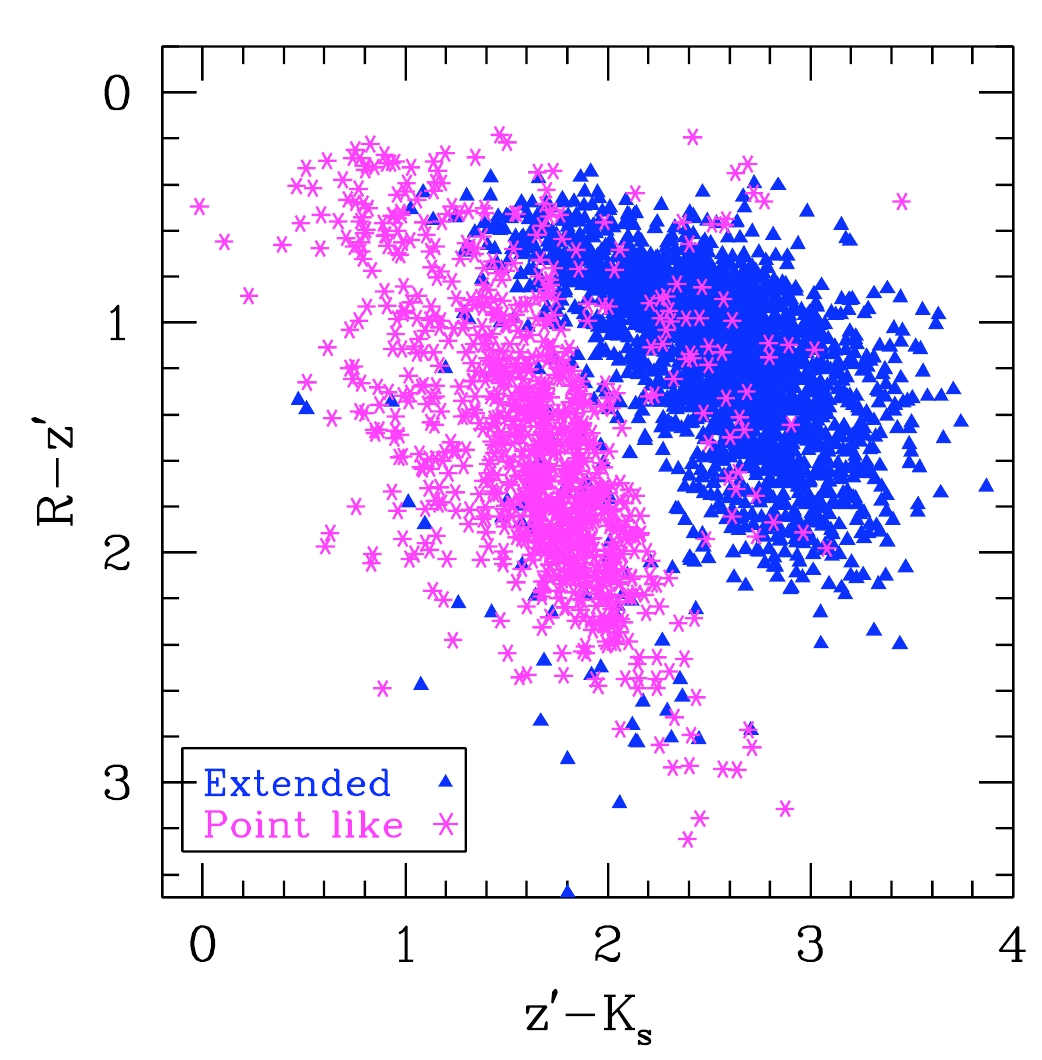}
\caption{(\rmz) vs (\zmk) diagram for the LCO point-like (asterisks)
and extended sources (triangles). The computed colors are
based on the 3$''$ photometry  (Vega magnitudes). The point-like
sources (mainly stars)  form a locus well separated from the region
occupied by the extended sources (galaxies).
For comparison we kept the axis convention of Fig.~2 from~\cite{warren00}, 
i.e. the vertical axis increases from top to bottom.}
\label{rmz_vs_zmK}
\end{figure}

The final matched catlogue features 3$''$ aperture photometry for all
bands where the photometry is computed using the magnitude of Vega as a
zero-point.  Fig.~\ref{rmz_vs_zmK} shows the (\rmz) vs (\zmk)
color-color diagram, where the majority of point-like and extended
sources occupy distinct regions.
Despite the clear separation of the two populations a small fraction
of point sources overlap with the extended ones and vice-versa,
possibly reflecting uncertainties in the photometry or the
morphological classification as well as intrinsically distinct
populations.
%
%
\section{\label{addata}Multi-wavelength data within the XMDS field}

This paper is primarily concerned with the properties of a quasar
sample selected using a variant of the $KX$ method, i.e. employing the
\rzk colors of optically point-like sources in the XMDS field.  In
order to discuss the physical nature of the $KX$-selected quasar
sample we will use photometric information at other wavelengths in
order to construct low resolution spectral energy distributions (SEDs)
for the candidate quasars.  In addition, in order to discuss any
possible bias in a $KX$-selected quasar sample, we will also compare
the $KX$-selected sample to quasar samples selected in the mid-infrared waveband.

We begin by providing a description of the X-ray data products
provided by the XMDS itself.  In addition, the CFHT Legacy Survey
(CFHTLS) and the Spitzer Wide-area InfraRed Extragalacic survey
(SWIRE) are two ``legacy''-type surveys that overlap the XMM Medium
Deep Survey (XMDS) field.  We provide below a brief description of the
data products associated with these surveys. Detailed information can
be found in~\cite{tajer07} and references therein.

The XMDS was conducted between July 2001 and January 2003 using the
European Photon Imaging Camera (EPIC) on board the XMM--Newton
satellite \citep{turner01,struder01}. It consists of 19 overlapping
pointings (of typical exposure time $20-25$ ksec) covering a
contiguous area of $\sim 2$ deg$^2$.  The energy range of the EPIC
instruments is between 0.1 to 12 keV.  A detailed description of the
pipeline used for reducing the X-ray observations and the catalog
properties can be found in~\cite{chiappetti05}.  The X-ray source
sample used in the present paper was generated from the
\cite{chiappetti05} source list by selecting unresolved sources
displaying more than 20 counts in the B ($0.5-2$) keV energy
band. This generated 370 X-ray sources in the \rzk catalog area.
 
The XMM-LSS survey (of which the XMDS is a subset) is coincident with
the Wide Synoptic component of the Canada France Hawaii Telescope
Legacy Survey1 (CFHTLS-W1)~\citep{gwyn07}.  CFHTLS-W1 $u^*, g', r',
i', z'$ photometry is available for approximately 90\% of the area
covered by the \rzk observations.  The magnitude limit of the survey,
at 50\% completeness, for a point source observed in seeing of
FWHM=0.8$''$ and a signal-to-noise ratio of 5 ($1.5\arcsec$ aperture
photometry), is 26.4, 26.6, 25.9, 25.5 and 24.8 for the $u^*,g',r',i'$
and $z'$ bands respectively (AB
magnitudes)\footnote{http://cfht.hawaii.edu/Science/CFHLS/cfhtlsgoals.html}.

SWIRE (\citealt{lonsdale03, lonsdale04}) is the largest Legacy Program
performed with the Spitzer Space Observatory~\citep{surace04}.  Once
again, 90\% of the sky area associated with the \rzk catalog is
covered by SWIRE data. The $5\sigma$ sensitivity for the 3.6, 4.5,
5.8, 8.0 and 24\micron\, energy bands are 7.3, 9.7, 27.5, 32.5 and 450
$\mu$Jy, respectively~\citep{lonsdale03}.  Throughout this work we
shall use the IRAC photometry measured within a 3$''$ aperture; for
the MIPS\,24\micron\, band the fluxes presented are those measured
using a 7.5$''$ diameter.

\section{A $KX$-selected quasar sample }
The original $KX$ method by~\cite{warren00} made use of the $V$, $J$
and $K$ bands, noting that quasars with a $V-J$ color similar to that
of stars would be redder in $J-K$, a method equally well suited
for the detection of reddened quasars according to the authors.  In
the present work we implement a variant of the original $KX$
technique, using the (\rmz) versus (\zmk) diagram instead.

\subsection{\label{stellar-locus}Seperating quasars from the stellar locus} 

\begin{figure}[h!]
\centering
\includegraphics[width=9cm]{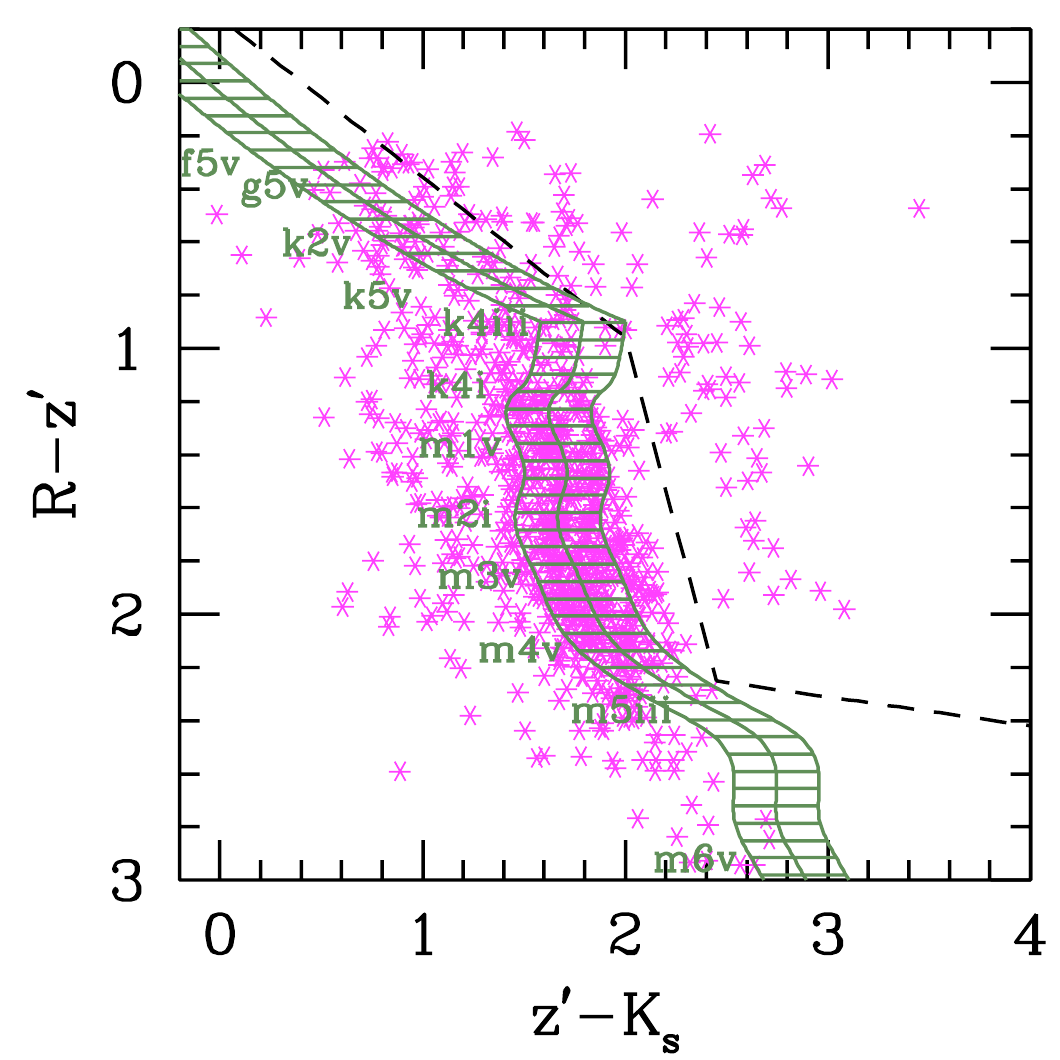}
\caption{Simulated stellar colors (shaded region) and distribution of
  the unresolved sources on the \rzk\ plane.  The dashed line
  indicates the adopted color threshold seperating stellar and
  non-stellar sources given by Equations 1-3.  }
\label{c-c-2dF-simul}.  
\end{figure}

Unresolved sources in the \rzk plane essentially consist of stars,
quasars and unresolved galaxies.  The properties of the stellar locus
on the \rzk plane were computed using the~\cite{pickles98} library of
130 stellar spectra (Fig.~\ref{c-c-2dF-simul}).  The locus of stellar
photometry was then extended along the \zmk\, axis to display a
half-width of $3\sigma$, where $\sigma$ is the photometric error on
the (\zmk) color index that dominates the photometric uncertainty in
the stellar locus (see also Table~\ref{general}).  Consideration of
the relative properties of the SEDs of stars and quasars indicates
that quasars should be located toward \zmk\, colors redder than the
stellar locus in Fig.~\ref{c-c-2dF-simul}.  We therefore identify
candidate quasars as unresolved sources displaying
%
\begin{equation}
	(R-z') = 0.6\,(z'-K_s) - 0.243 \hspace{0.6cm}\mathrm{for}\hspace{0.2cm} (z'-K_s) < 2.0,
\end{equation}
%
\begin{equation}
	(R-z') = 2.877\,(z'-K_s) - 4.797 \hspace{0.2cm}\mathrm{for}\hspace{0.2cm} 2.0 \le (z'-K_s) \le 2.4,
\end{equation}
%
%
\begin{equation}
	(R-z') = 0.109\,(z'-K_s) + 1.983 \hspace{0.6cm}\mathrm{for}\hspace{0.2cm} (z'-K_s) > 2.4.
\end{equation}
%
These selection criteria identify 93 candidate quasars.
%

\subsection{\label{qso-identif}SED analysis of quasar candidates}

In the absence of low- to medium-resolution spectroscopy we
investigate the nature of the candidate quasar sample via analysis of
their UV to MIR SEDs. The \rzk catalog was therefore matched to the
CFHTLS and SWIRE catalogs in order to reconstruct the SED of the
objects.  The CFHTLS catalog was matched to the \rzk detections using
a matching radius of 2$''$.  Thanks to the large overlap between the
two sets of observations, CFHTLS $u^*,g',r',i',z'$ counterparts were
found for 86/93 quasar candidates.
Furthermore, 90/93 quasar candidates are matched to SWIRE sources
using a tolerance of $1.5\arcsec$ and show emission in the first two
IRAC bands (IRAC\,1 = 3.6\micron\,, IRAC\,2 =
4.5\,\micron~\citep{fazio04}). At longer wavelengths, however, the
number decreases significantly with $\sim 30$\% of the candidates
showing emission in the IRAC\,3 (5.8\micron) and IRAC\,4 (8.0\micron)
band and only 16\% with emission in the MIPS~24\micron\, channel
\citep{rieke04}.
Therefore, the SED of the quasar candidates will include up to
13 photometric points ($u^*,g',r',R,i',z',z'_{ctio}, K_s$,
IRAC\,1,\,2,\,3\,\,4 and MIPS\,24 bands), spanning the range $0.4-24\,
\mu m$.

\begin{figure*}
\centering
\includegraphics[width=16.0cm]{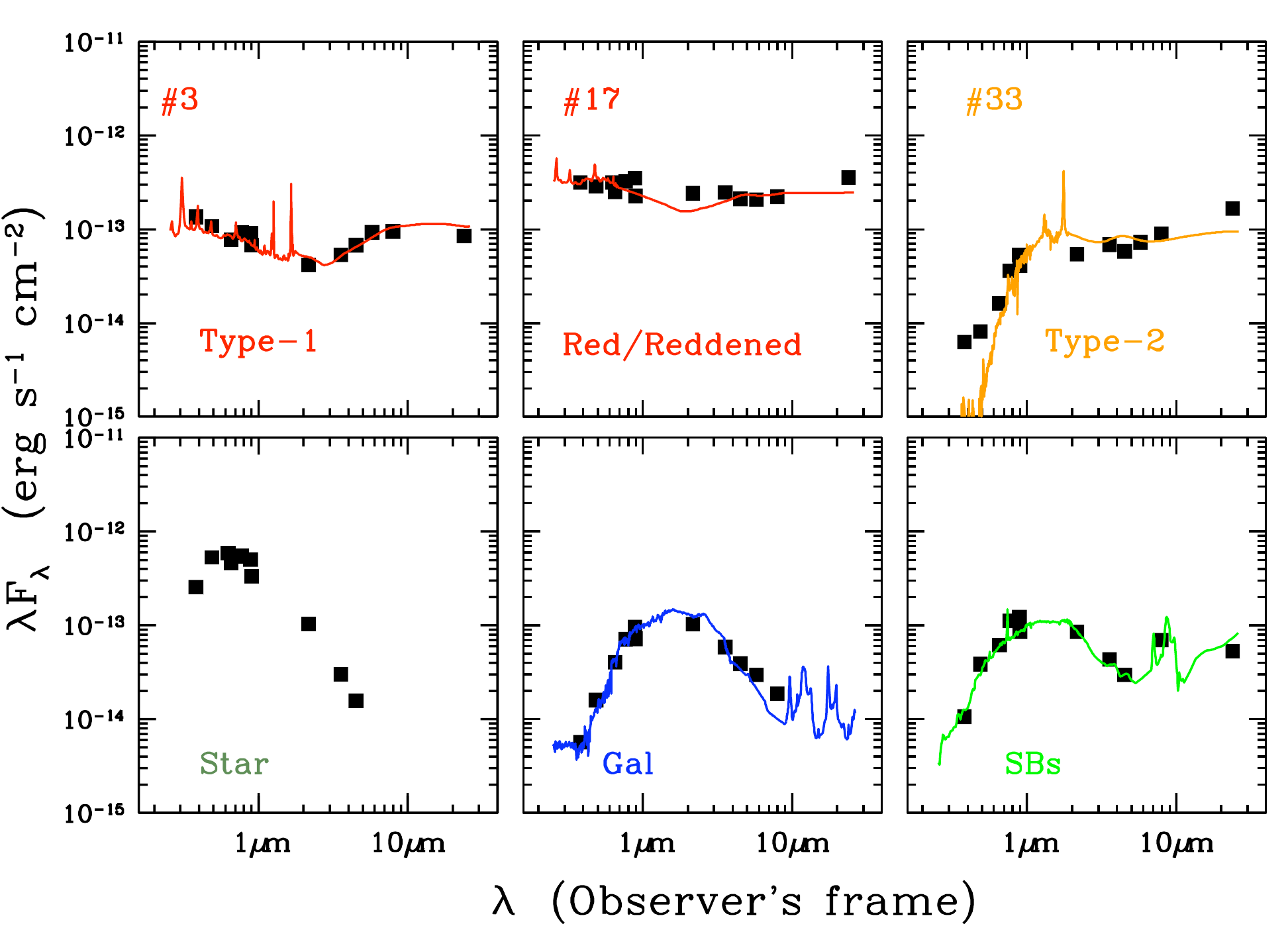}
\caption{{\it Upper Panel:} Typical examples of the Spectral Energy
  Distribution of: a ``blue" type-1 quasar (left), a quasar showing
  significant reddening signature (middle) and of the single type-2
  quasar found in the LCO survey (right).  In each case the best-fit
  template (as derived from the HyperZ calculation \---\ see
  Section~\ref{photoz}) is also plotted.  Objects \#3, \#17 and \#33
  are found at redshifts of 1.513, 0.710 and 1.679, respectively.
  {\it Lower Panel}: SED of a star (left), an elliptical galaxy at
  redshift $z = 0.545$ (middle) and a galaxy showing starburst
  activity at $z = $0.127. }
\label{sed-examples}
\end{figure*}

The investigation of the candidate quasars sample was performed
initially via a visual inspection of their SEDs. Quasar SEDs exhibit
two unmistakable signatures in the UV/optical/IR domain: i) the
UV/optical-bump customarily attributed to the accretion process; and
ii) the minimum (in $\lambda F_{\lambda}$) or break with subsequent
change of slope (in $F_{\lambda}$ or $F_{\nu}$), occurring at around 1
micron (see \citealt{hatzi05}; \citealt{richards06}), where the drop
of the accretion bump meets with the T $\sim 1500$ K black-body rise
(towards the IR), corresponding to the graphite grains believed to be
one of the two main components of the dusty tori surrounding the
active nucleus.  This visual inspection yielded 25 quasars among the 93
candidates.
The remaining sources were characterised as late type stars and (low
redshift) galaxies, characterised either as early-type or
starburst-type.  The SED of elliptical galaxies peaks around 2 to 3
micron and then decreases monotonically towards redder wavelengths,
while that of starbursts shows an excess in the IRAC\,4 band due to
the strong PAH emission centered at 7.7 micron~\citep{weedman06}.
For the quasars, 22/25 ($\sim 82$\%) are ``classical''
(i.e. blue) type-1, and one ($\sim 6$\%) is of type-2 (as its flux
experiences a significant attenuation at bluer wavelengths).
Although, at present, there is no clear definition of a red quasar
(\citealt{canalizo06} and references therein), there is no doubt that
there exists a fraction of quasars where the optical spectrum deviates
significantly from that of a type-1 object (e.g.~\citealt{pierre01};
\citealt{leipski07b}).  Two entries in our quasar sample belong to
this category, showing a signature of moderate reddening with a flat
SED blue-ward of the $\sim$1\micron\ inflection.  The lack of
spectroscopic information and the small number of objects do not allow 
any firm conclusion regarding the origin of the flattening of
their optical SED.  A possible explanation could be the presence of a
moderate amount of dust along the line of sight to the active
nucleus. In this scenario the dust would affect differentially the
blue and red light emitted by the quasar, making the SED appear
flatter compared to that of a blue type-1 quasar, but not as steep as
that of a type-2. However, larger numbers of sources displaying this
feature would be needed before claiming any robust conclusion.
The results of the SED inspection are presented in Table~\ref{sed-res}
and examples of visually classified SEDs are shown in
Fig.~\ref{sed-examples}.
%
\begin{table}
\caption{Results of the SED inspection for the 93 quasar candidates.}
\label{sed-res}
\centering
\begin{tabular}{|l|c|}
\hline
  SED type            & Number \\
                              \hline
                              \hline
QSO Type-1            &	   22 \\
QSO Type-1 (reddened) &	    2 \\
QSO Type-2	          &	    1 \\
Galaxy                &     50 \\
Star                  & 18 \\
   \hline
\end{tabular}
\end{table}
%

%
%

\subsection{\label{mir}The MIR color-color diagram}

\begin{figure}
\centering
\includegraphics[width=8.0cm]{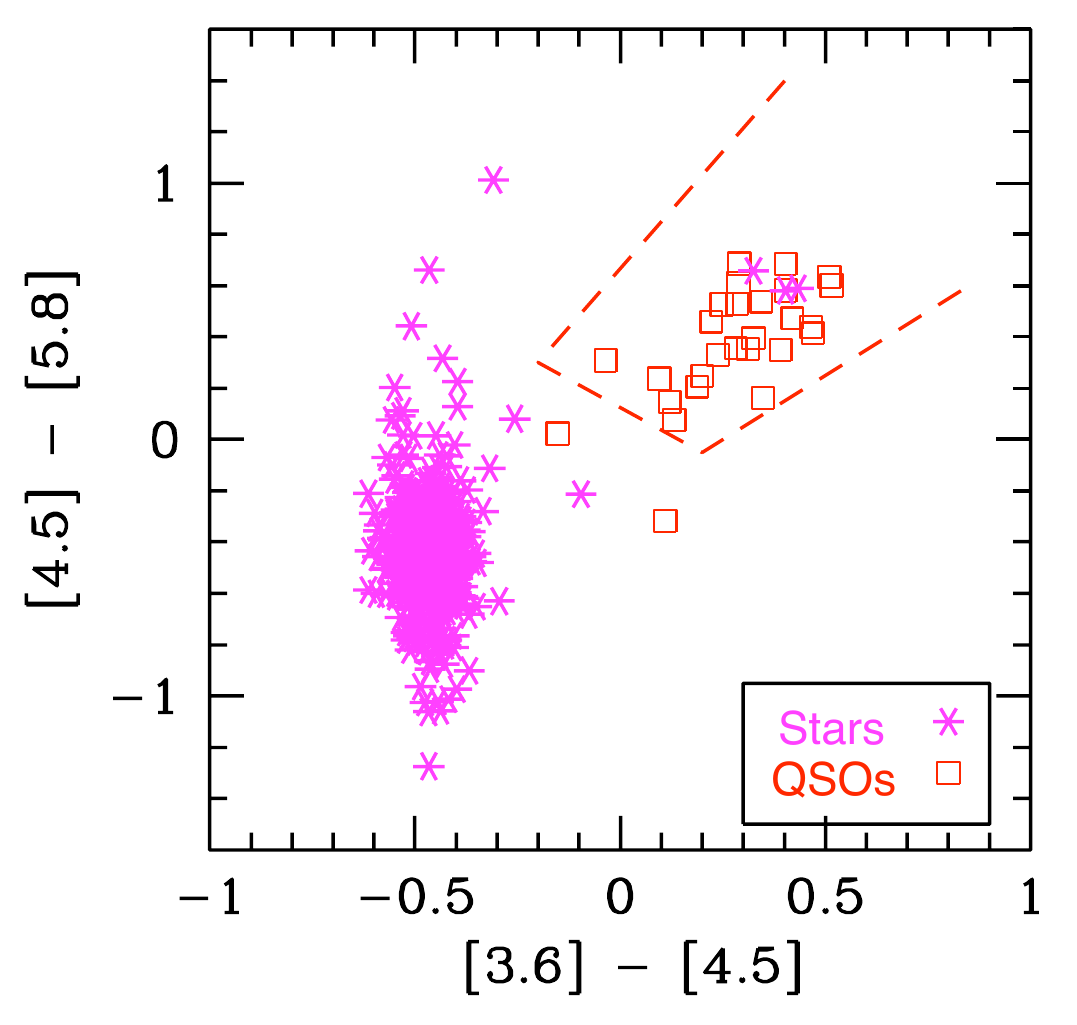}
\caption{SWIRE color-color diagram for unresolved sources in the \rzk
  catalog. Sources classified as stars are displayed as asterisks and
  sources classified as quasars (on the basis of \rzk colors and SED
  inspection) are plotted as open squares.  According
  to~\cite{hatzi05} quasars are expected to be found in the region
  delineated with the long-dashed line.  Visual inspection of the SED
  of the three ``stars'' within the quasar locus confirms the quasar
  hypothesis.}
\label{swire-colors}
\end{figure}

The $KX$ method seperates quasars and stars on the basis of their
optical and near-infrared colors. An alternative method applies the
same test on the mid-infrared colour plane.  \cite{hatzi05}
demonstrated that it is possible to isolate type-1 quasars using the
first three IRAC bands (3.6, 4.5 and $5.8\,\mu m)$.  These bands
seperate quasars from stars as they sample the stellar Rayleigh-Jeans
tail of the blackbody spectrum. As a result, the stellar colors are
very close to zero in the Vega photometric system.  Following their
approach, we constructed the MIR color-color diagram for all
unresolved sources in the \rzk catalog
(Fig.~\ref{swire-colors}). Three sources classified as stars on the
basis of their \rzk photometry are located in the so-called ``quasar
locus'' and visual inspection of their SEDs confirms the quasar
hypothesis. Of these sources, two are located very close (about 0.2
mag in each direction) to the stellar locus on the \rzk plane. The
third source is an isolated quasar located in the stellar locus of
Fig.~\ref{rmz_vs_zmK}, for which we possess a spectroscopic
identification (see section~\ref{2df-spectra}).  We finally note that
two sources classified as quasars on the basis of their \rzk
photometry and SED analysis lie outwith yet close to the quasar locus
defined by \cite{hatzi05}.

%

\subsection{\label{2df-spectra} Spectroscopy of quasar candidates} 

\begin{figure} 
\centering
\includegraphics[width=9cm]{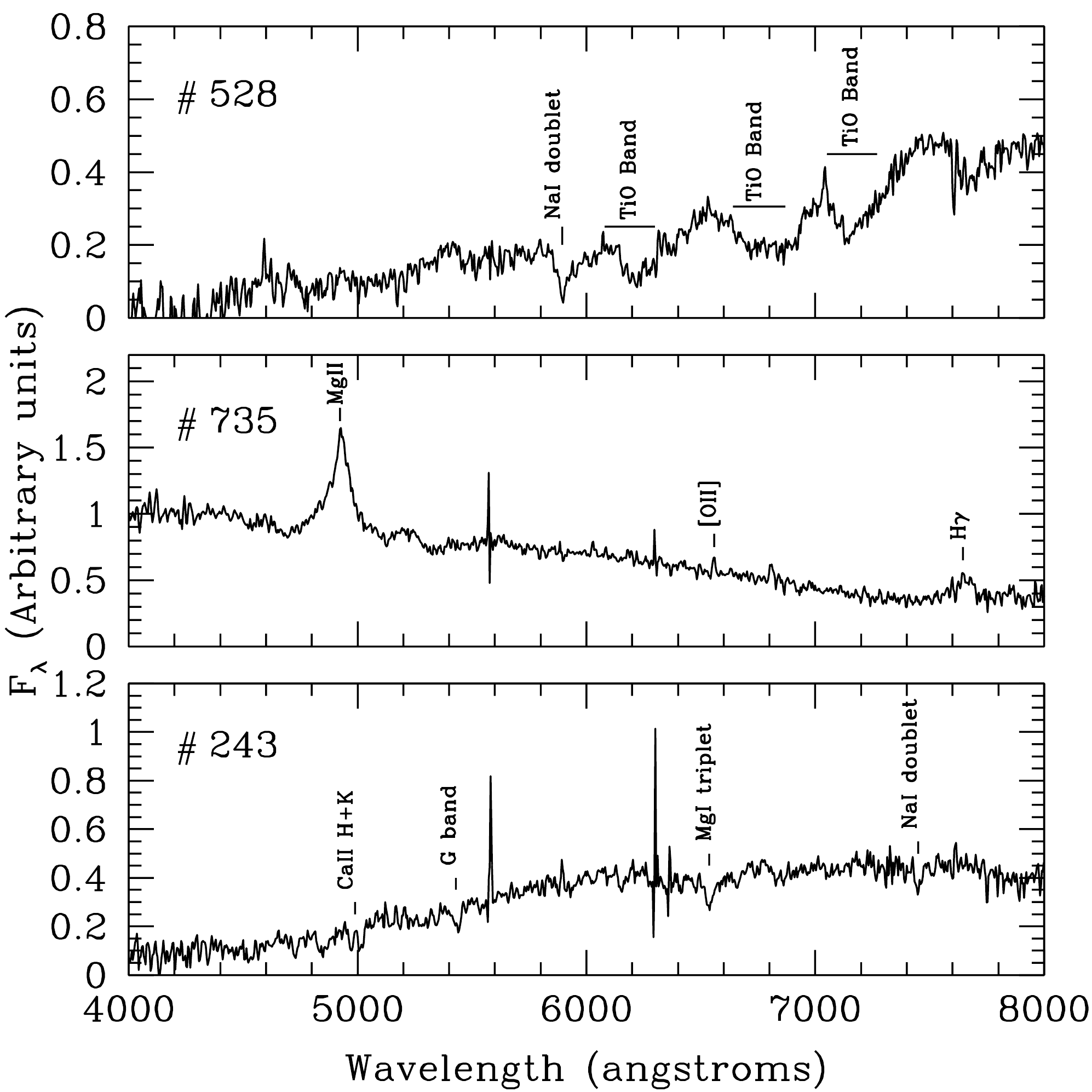} 
\caption{Examples of good-quality 2dF spectra used to investigate \rzk
  sources. The identification of each object is presented in the upper
  left part of each spectrum; some of the spectral features employed
  to classify each source are also shown. Object \#\,528 is a star, object
  \#\,735 a quasar at $z = 0.76$ and object \#\,243 a galaxy at $z =
  0.263$~\citep{garcet07}.}
\label{spec-ex}
\end{figure}

To date it has not been possible to perform a systematic spectroscopic
investigation of the $KX$-selected quasar sample presented in this
paper. However, a number of spectroscopic programs have been performed
which overlap the XMDS area with the result that spectroscopic information is
available for a subset of sources presented in this paper. As part of
the spectroscopic identification program for the XMM-Newton Survey Science 
Centre medium sensitivity serendipitous 
survey~\cite{tedds06}, the two degree field multi object spectrograph~(2dF, 
\citealt{lewis02})
targeted some 800 X-ray detected sources with $r_{AB} < 22$ including 
the XMM-LSS area.  More details about the 2dF program will be given in a 
forthcoming X-ray Wide Area Survey (XWAS) catalog paper (Tedds et al. in prep.).
A total of 22 sources with 2dF spectra were associated with sources in
the \rzk catalog.  These included one star, 9 quasars and 12 galaxies
as classified by their \rzk photometry.  In each case the
spectroscopic classification matched the \rzk classification.  A
further five spectra were available for sources with $K_s>18$ and
therefore not included in the \rzk catalog. These sources were
classified as type-1 quasars. They are not used further in the
analysis, except as part of a spectral training set for the
photometric redshift analysis (Section~\ref{photoz}).  Three examples
of good-quality 2dF spectra are presented in Fig.~\ref{spec-ex}.
%
%

A further four sources classified as quasars on the basis of their
\rzk photometry and SED analysis have been confirmed spectroscopically
as type-1 quasars (source numbers 2, 3, 20 and 27 in Table
2). Observations were conducted as part of the VLT/VIMOS program
(080.A-0852B) to investigate X-ray selected AGN in the XMM-LSS survey
area (Garcet et al. in prep.).

Therefore, for the $KX$-selected quasar sample, 12/25 sources classified
as quasars on the basis of \rzk photometry and SED analysis have been
confirmed spectroscopically. However, it is important to note that the
spectroscopic programs that observed these sources were constructed to
investigate X-ray selected source samples. It is therefore important
at this stage to consider the X-ray properties of the $KX$-selected
quasar sample.

\subsection{\label{x-rays}X-ray properties of the quasar sample}

As noted in Section~\ref{addata}, there are 370 X-ray sources in the
overlapping XMDS and \rzk catalog area. Of these, 81 X-ray sources
were matched to a single \rzk source within a $7\arcsec$ matching
radius. This radius was chosen on the basis of: (a) the density of
X-ray detections; (b) the accuracy of the X-ray astrometric solution
($\sim1''$); and (c) XMM-Newton's point spread function ($6\arcsec$).
 
The relatively low fraction of matched sources, 81/370,
is largely due to the shallow depth of the \ks\ band observations
compared to the optical observations. Approximately 20\% of the
$KX$-selected quasar sample, 21/93, are matched to an X-ray
detection. Of these matched sources, 12/13 are spectroscopically
confirmed quasars and 21/25 are SED classified quasars (note that the
SED classified quasars include all of the spectroscopically confirmed
quasars as well). None of the sources classified as stars or galaxies
as a result of the SED inspection have an X-ray counterpart.

\begin{figure} 
\centering
\includegraphics[width=9cm]{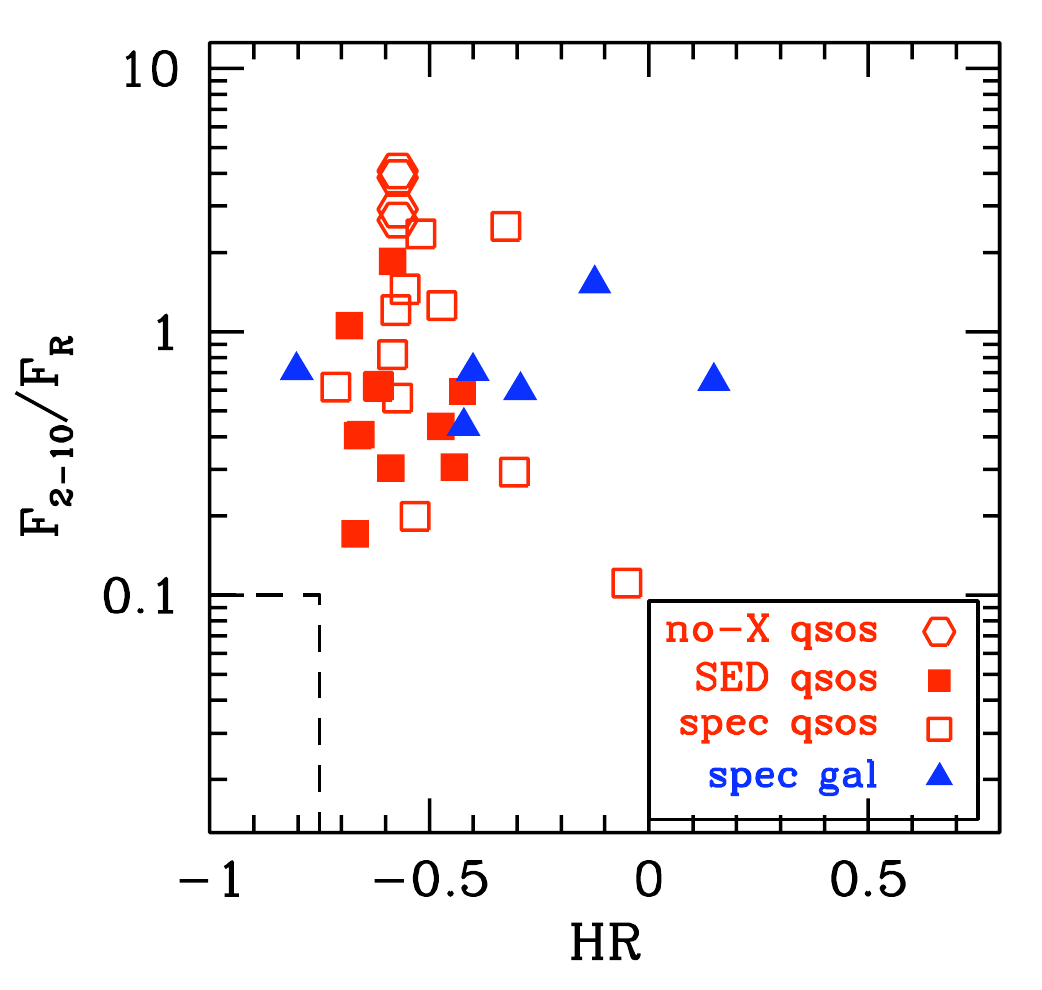} 
\caption{Plot of the X-ray to optical flux ratio for SED and spectroscopically
  classified quasars and spectroscopically classified galaxies (from
  the 2dF sample) versus X-ray hardness ratio. Hardness ratio is
  computed according to Equation 2 of ~\cite{garcet07}. The four quasar
  candidates that are undetected in X-rays are plotted as upper limits
  on $\rm F_{2-10}/F_R$ with the median HR value of the X-ray detected
  quasar sample. The region enclosed by the dashed rectangle indicates
  the anticipated locus of X-ray emitting stars.}
\label{ratio_xopt}
\end{figure}

We intend to discuss the X-ray properties of the $KX$-selected quasar
sample in greater detail in a subsequent paper.  However, at this
point we note that the X-ray-to-optical flux ratios of the
SED-classified quasars occupy the interval \mbox{$0.1 < \log(F_X/F_R)
  < 10.0$}, typically occupied by AGN~\citep{ceca04}.
Figure~\ref{ratio_xopt} indicates that the four sources classified as
quasars yet lacking an X-ray counterpart are consistent with being
drawn from the same population as the X-ray detected quasars if we
consider that application of the above X-ray-to-optical flux ratio
would place the X-ray fluxes for these four sources below the nominal
flux limit of the XMDS X-ray source catalog.

Therefore, for the sample of 93 $KX$-selected quasars, 21/25 of the
sources classified as quasars from the SED analysis show X-ray
emission. The remaining four sources are consistent with being drawn
from the same population. None of the 68 remaining sources classified
as either stars or galaxies from the SED analysis is detected in X-rays. 
We therefore conclude that the X-ray properties of the
$KX$-selected quasar sample support the previous conclusions on the
nature of individual sources derived from the SED analysis.

\subsection{\label{photoz}Photometric redshift analysis}

In the absence of complete spectroscopic informtion for the 93 sources
in the $KX$-selected quasar sample we employed the HyperZ
code~\citep{bolzon00} to compute photometric redshifts.  The spectral
templates applied to the data are described in~\cite{polletta07} and
cover the wavelength interval 100 nm $-$ 100 $\mu$m.  We tested the
performance of HyperZ by first running the program on the 14 quasars
with 2dF spectra, using various combinations of filters, reddening
laws and templates.  Based on the limiting magnitude of the \rzk
survey and the maximum spectroscopic redshift in our catalog, we set
$z_{max} = 3.0$. As HyperZ is limited to 15 templates per run, we
selected five quasar templates (three type-1 and two type-2) while the
remainder were selected among the templates for elliptical, starburst,
Seyfert and spiral galaxies.

The performance of the photometric redshift method was estimated on
the basis of two parameters: the fractional error $\Delta z$ and the
$1 \sigma$ dispersion $\sigma_z$ as defined by~\cite{tajer07},
\begin{equation}
	\Delta z = \left(\frac{z_{phot}-z_{spec}}{1+z_{spec}}\right)
\hspace{0.5cm}\mathrm{and}\hspace{0.5cm}
	\sigma_z^2 = \frac{1}{N} \sum \left(\frac{z_{phot}-z_{spec}}{1+z_{spec}}\right)^2
\end{equation}
where $N$ is the number of sources with spectroscopic redshifts.  After
various trials we obtained a photometric redshift for 12 of the 14
sources with $\left| \Delta z \right| < 0.3$ and $\sigma_z = 0.19$.

\begin{figure*}[t]
\centering
\includegraphics[width=12.0cm]{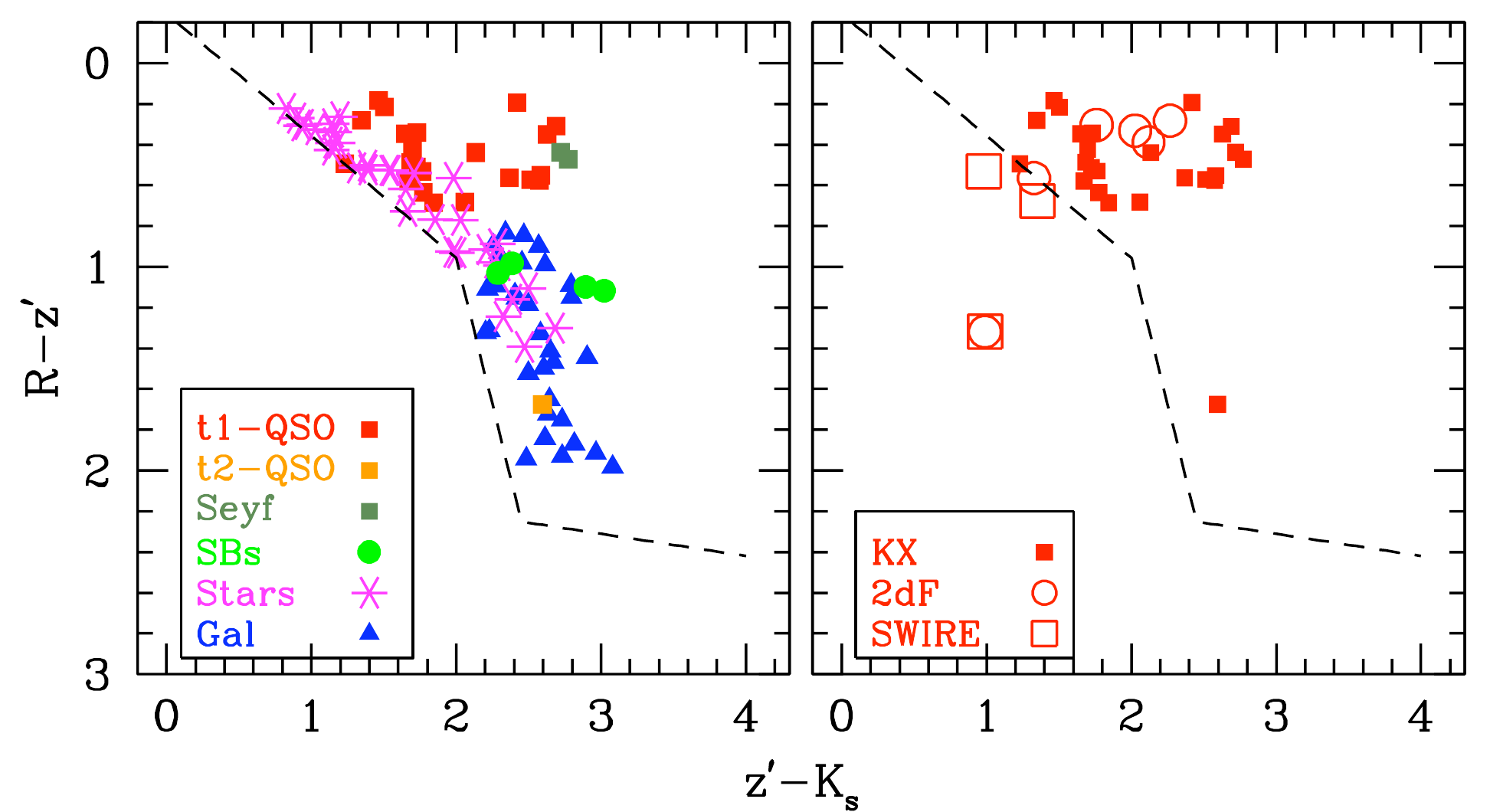}
\caption{ {\it Left panel:} Distribution in the color-color diagram of
  the initial sample of 93 quasar candidates. The color-coding is
  based on the best-fitting SED returned via the photometric redshift
  analysis. The dashed line indicates the border between the stellar
  and non-stellar colors used in our selection procedure.  {\it Right
    panel:} Color-color diagram containing the quasars detected using
  jointly several methods (2dF, $KX$, SWIRE).  The $KX$-selected
  quasars are marked with filled squares while the open circles denote
  the 6 quasars with 2dF spectra not selected by the $KX$.  Three
  type-1 quasars, selected on the basis of their IRAC colors, are
  marked as open squares. The isolated object at $(R-z')\approx 1.7$
  is the type-2 $KX$-selected quasar found in our survey.}
\label{final-c-c}
\end{figure*}

The remaining 16 quasars without spectra consists of 15 type-1 and one
type-2 object (\#33), characterized as such on the basis of their SED.
The best-fit template matched by HyperZ was in very good agreement
with the SED-based classification, fitting the ID\,\#33 quasar with a
type-2 spectral template and 13 out of 15 type-1 objects with a type-1
spectral template. Only two type-1 objects were fit by a Seyfert 1.8
template (\#27 and \#57).  Figure~\ref{final-c-c} shows the template
fitting results for the full sample of 93 quasar candidates. Although
the upper limit for the $A_V$ extinction coefficient was set to 0.5
no object was found to have $A_V > 0.4$, with 13 out of 19 objects
having $A_V < 0.2$.  Finally, the $B$-band absolute magnitude was set
to vary between $ -23.0 < M_B < -28.8$, but no object was found to
have an absolute magnitude brighter than $M_B = -26.22$.  The final
photometric redshifts for the 16 newly discovered quasars are
presented in Table~\ref{type1-type2-qso-p1}.

%
%

\section{\label{disc}Conclusions}

The distribution in \rzk for the 93 quasar candidates is shown in
the left panel of Fig.~\ref{final-c-c}. The distribution of all
quasars found in the \rzk catalog, using $KX$-selection plus SED
analysis, 2dF spectra (subsection~\ref{2df-spectra}) and IRAC colors
(subsection~\ref{mir}) is presented in the right panel of the same
figure. The first conclusion to be drawn from these two plots is that
the majority of the quasar population (type-1) is concentrated in the
bluer part of the (\rmz) axis, with the (\rmz) color ranging from 0.2
to 0.8 magnitudes. This occurs as the quasar power-law SED produces an
excess in the \ks-band, with the result that quasars appear redder
than stars along the \zmk\ axis.  In addition, the same power-law SED
causes quasars to appear bluer than galaxies.  We note the presence of
a single type-2 quasar detected by the $KX$-method. This object is
much redder, with \rmz $\approx 1.7$, and would have been missed if
stricter color criteria had been applied, e.g. a bluer cut in \rmz.

To assess the performance of the $KX$-method in a quantitative manner
we adopt the formalism presented in~\cite{hatzi00}. Defining $N_c$ as
the number of quasar selected candidates, $N_f$ as the number of
quasars confirmed from the candidates and $N_e$ as the number of
expected quasars (based upon model predictions), one may define the
efficiency, $E$, of the technique as
\begin{equation}
{
E = \frac{N_f^2}{N_e N_c}.
}
\label{eqn-effic}
\end{equation}
The $KX$-method identified 25 quasars in an area of 0.68\,deg$^2$ at
$K_s < 18$. The \rzk catalog is 80\%\ complete at $K_s=18$ and
therefore the effective surface density of $KX$-selected quasars at
$K_s<18$ is $46 \pm 9 \deg^{-2}$. \cite{maddox06} use model
predictions to estimate a quasar surface density at $K_s<18$ of $50
\deg^{-2}$.  Employing these values, the completeness of the
$KX$-selected sample is on the order of 90\%.  The identification of 25
quasars from the 93 candidates indicates a confirmation rate on the order
of 27\%.  Therefore, employing Equation~\ref{eqn-effic}, the overall
efficiency of the $KX$-techniqe as applied in this paper is
approximately 25\%.

Detailed studies of quasar populations represent a key component of
the current generation of wide-field optical to MIR imaging surveys.
Selection by color provides an important technique for compiling large
samples of quasar candidates.  However, the application of color
criteria alone are unlikely to generate uncontaminated quasar samples.
SED classification of quasar candidates may provide an important
technique for prioritizing the spectroscopic observations of large
quasar samples \---\ with the potential to improve considerably the
efficiency of follow-up programs.  Throughout the present analysis we
have applied a visual classification of candidate quasar SEDs.
Clearly an automated approach will be required for larger samples.
Large surface IR surveys, such as UKIDSS~\citep{hewett06}, will base
their quasar selection on the $KX$ technique. Because of the depth of
the surveys and the large area coverage (several thousand square
degrees) it would be advantageous to minimize contamination prior to
spectroscopic confirmation.

\begin{acknowledgements}
This paper is based on observations obtained at Cerro Tololo Inter-American 
Observatory a division of the National Optical Astronomy Observatories, 
which is operated by the Association of Universities for Research in 
Astronomy, Inc. under cooperative agreement with the National Science Foundation.
It also includes data obtained with the 2.5 meter Du Pont telescope 
located at Las Campanas Observatory, Chile.\\
This publication makes use of data products from the Two Micron All Sky Survey, 
which is a joint project of the University of Massachusetts and the Infrared Processing 
and Analysis Center/California Institute of Technology, funded by the National 
Aeronautics and Space Administration and the National Science Foundation.\\
This work is partially based on observations made with the Spitzer Space Telescope, 
which is operated by the Jet Propulsion Laboratory, California Institute of Technology 
under a contract with NASA. \\
Based on observations obtained with MegaPrime/MegaCam, a joint project of 
CFHT and CEA/DAPNIA, at the Canada-France-Hawaii Telescope (CFHT) which is 
operated by the National Research Council (NRC) of Canada, the Institut National 
des Science de l'Univers of the Centre National de la Recherche Scientifique 
(CNRS) of France, and the University of Hawaii. This work is based in part 
on data products produced at TERAPIX and the Canadian Astronomy Data 
Centre as part of the Canada-France-Hawaii Telescope Legacy Survey, 
a collaborative project of NRC and CNRS.\\
Th.N. wishes to thank M. Polletta (IAP) for  fruitful
discussions concerning the photometric redshifts. He also wishes to thank
A. Smette (ESO) and P. Gandhi (RIKEN) for advice at the data reduction step.
Th.N., P.R., J.S. and O.G. acknowledge the ESA PRODEX 
Programme ``XMM-LSS" and the Belgian Federal Science Policy 
Office for their support. GG and HQ thank FONDAP 15010003 Center for Astrophysics.
JPW acknowledges the support of the Natural Sciences and Engineering Research 
Council of Canada (NSERC).
MJP, JAT and SM acknowledge support from the UK PPARC and STFC research councils. The 2df spectrograph was mounted on the Anglo Australian Telescope and funded by the UK and Australian research councils.
\end{acknowledgements}

\bibliographystyle{aa}
\bibliography{9584}

\begin{thebibliography}{55}
\expandafter\ifx\csname natexlab\endcsname\relax\def\natexlab#1{#1}\fi

\bibitem[{{Andreon} {et~al.}(2004){Andreon}, {Willis}, {Quintana},
  {Valtchanov}, {Pierre}, \& {Pacaud}}]{andreon04}
{Andreon}, S., {Willis}, J., {Quintana}, H., {et~al.} 2004, \mnras, 353, 353

\bibitem[{{Barkhouse} \& {Hall}(2001)}]{barkhouse01}
{Barkhouse}, W.~A. \& {Hall}, P.~B. 2001, \aj, 121, 2843

\bibitem[{{Benn} {et~al.}(1998){Benn}, {Vigotti}, {Carballo},
  {Gonzalez-Serrano}, \& {S{\'a}nchez}}]{benn98}
{Benn}, C.~R., {Vigotti}, M., {Carballo}, R., {Gonzalez-Serrano}, J.~I., \&
  {S{\'a}nchez}, S.~F. 1998, \mnras, 295, 451

\bibitem[{{Bertin} \& {Arnouts}(1996)}]{bertin96}
{Bertin}, E. \& {Arnouts}, S. 1996, \aaps, 117, 393

\bibitem[{{Bolzonella} {et~al.}(2000){Bolzonella}, {Miralles}, \&
  {Pell{\'o}}}]{bolzon00}
{Bolzonella}, M., {Miralles}, J.-M., \& {Pell{\'o}}, R. 2000, \aap, 363, 476

\bibitem[{{Brown} {et~al.}(2006){Brown}, {Brand}, {Dey}, {Jannuzi}, {Cool}, {Le
  Floc'h}, {Kochanek}, {Armus}, {Bian}, {Higdon}, {Higdon}, {Papovich},
  {Rieke}, {Rieke}, {Smith}, {Soifer}, \& {Weedman}}]{brown06}
{Brown}, M.~J.~I., {Brand}, K., {Dey}, A., {et~al.} 2006, \apj, 638, 88

\bibitem[{{Calzetti} {et~al.}(2000){Calzetti}, {Armus}, {Bohlin}, {Kinney},
  {Koornneef}, \& {Storchi-Bergmann}}]{calzetti00}
{Calzetti}, D., {Armus}, L., {Bohlin}, R.~C., {et~al.} 2000, \apj, 533, 682

\bibitem[{{Canalizo} {et~al.}(2006){Canalizo}, {Stockton}, {Brotherton}, \&
  {Lacy}}]{canalizo06}
{Canalizo}, G., {Stockton}, A., {Brotherton}, M.~S., \& {Lacy}, M. 2006, New
  Astronomy Review, 50, 650

\bibitem[{{Chen} {et~al.}(2002){Chen}, {McCarthy}, {Marzke}, {Wilson},
  {Carlberg}, {Firth}, {Persson}, {Sabbey}, {Lewis}, {McMahon}, {Lahav},
  {Ellis}, {Martini}, {Abraham}, {Oemler}, {Murphy}, {Somerville}, {Beckett},
  \& {Mackay}}]{chen02}
{Chen}, H.-W., {McCarthy}, P.~J., {Marzke}, R.~O., {et~al.} 2002, \apj, 570, 54

\bibitem[{{Chiappetti} {et~al.}(2005){Chiappetti}, {Tajer}, {Trinchieri},
  {Maccagni}, {Maraschi}, {Paioro}, {Pierre}, \& {Surdej}}]{chiappetti05}
{Chiappetti}, L., {Tajer}, M., {Trinchieri}, G., {et~al.} 2005, \aap, 439, 413

\bibitem[{{Croom} {et~al.}(2001){Croom}, {Warren}, \& {Glazebrook}}]{croom01}
{Croom}, S.~M., {Warren}, S.~J., \& {Glazebrook}, K. 2001, \mnras, 328, 150

\bibitem[{{Della Ceca} {et~al.}(2004){Della Ceca}, {Maccacaro}, {Caccianiga},
  {Severgnini}, {Braito}, {Barcons}, {Carrera}, {Watson}, {Tedds}, {Brunner},
  {Lehmann}, {Page}, {Lamer}, \& {Schwope}}]{ceca04}
{Della Ceca}, R., {Maccacaro}, T., {Caccianiga}, A., {et~al.} 2004, \aap, 428,
  383

\bibitem[{{Fazio} {et~al.}(2004){Fazio}, {Hora}, {Allen}, {Ashby}, {Barmby},
  {Deutsch}, {Huang}, {Kleiner}, {Marengo}, {Megeath}, {Melnick}, {Pahre},
  {Patten}, {Polizotti}, {Smith}, {Taylor}, {Wang}, {Willner}, {Hoffmann},
  {Pipher}, {Forrest}, {McMurty}, {McCreight}, {McKelvey}, {McMurray}, {Koch},
  {Moseley}, {Arendt}, {Mentzell}, {Marx}, {Losch}, {Mayman}, {Eichhorn},
  {Krebs}, {Jhabvala}, {Gezari}, {Fixsen}, {Flores}, {Shakoorzadeh}, {Jungo},
  {Hakun}, {Workman}, {Karpati}, {Kichak}, {Whitley}, {Mann}, {Tollestrup},
  {Eisenhardt}, {Stern}, {Gorjian}, {Bhattacharya}, {Carey}, {Nelson},
  {Glaccum}, {Lacy}, {Lowrance}, {Laine}, {Reach}, {Stauffer}, {Surace},
  {Wilson}, {Wright}, {Hoffman}, {Domingo}, \& {Cohen}}]{fazio04}
{Fazio}, G.~G., {Hora}, J.~L., {Allen}, L.~E., {et~al.} 2004, \apjs, 154, 10

\bibitem[{{Francis} {et~al.}(2000){Francis}, {Whiting}, \&
  {Webster}}]{francis00}
{Francis}, P.~J., {Whiting}, M.~T., \& {Webster}, R.~L. 2000, \pasa, 17, 56

\bibitem[{{Garcet} {et~al.}(2007){Garcet}, {Gandhi}, {Gosset}, {Sprimont},
  {Surdej}, {Borkowski}, {Tajer}, {Pacaud}, {Pierre}, {Chiappetti}, {Maccagni},
  {Page}, {Carrera}, {Tedds}, {Mateos}, {Krumpe}, {Contini}, {Corral},
  {Ebrero}, {Gavignaud}, {Schwope}, {Le F{\`e}vre}, {Polletta}, {Rosen},
  {Lonsdale}, {Watson}, {Borczyk}, \& {Vaisanen}}]{garcet07}
{Garcet}, O., {Gandhi}, P., {Gosset}, E., {et~al.} 2007, \aap, 474, 473

\bibitem[{{Gwyn}(2007)}]{gwyn07}
{Gwyn}, S.~D.~J. 2007, ArXiv e-prints, 710

\bibitem[{Hall {et~al.}(1998)Hall, Green, \& Cohen}]{hall98}
Hall, P., Green, R., \& Cohen, M. 1998, \apjs, 119, 999

\bibitem[{{Hatziminaoglou} {et~al.}(2000){Hatziminaoglou}, {Mathez}, \&
  {Pell{\'o}}}]{hatzi00}
{Hatziminaoglou}, E., {Mathez}, G., \& {Pell{\'o}}, R. 2000, \aap, 359, 9

\bibitem[{{Hatziminaoglou} {et~al.}(2005){Hatziminaoglou}, {P{\'e}rez-Fournon},
  {Polletta}, {Afonso-Luis}, {Hern{\'a}n-Caballero}, {Montenegro-Montes},
  {Lonsdale}, {Xu}, {Franceschini}, {Rowan-Robinson}, {Babbedge}, {Smith},
  {Surace}, {Shupe}, {Fang}, {Farrah}, {Oliver}, {Gonz{\'a}lez-Solares}, \&
  {Serjeant}}]{hatzi05}
{Hatziminaoglou}, E., {P{\'e}rez-Fournon}, I., {Polletta}, M., {et~al.} 2005,
  \aj, 129, 1198

\bibitem[{{Hewett} {et~al.}(2006){Hewett}, {Warren}, {Leggett}, \&
  {Hodgkin}}]{hewett06}
{Hewett}, P.~C., {Warren}, S.~J., {Leggett}, S.~K., \& {Hodgkin}, S.~T. 2006,
  \mnras, 367, 454

\bibitem[{{Jurek} {et~al.}(2007){Jurek}, {Drinkwater}, {Francis}, \&
  {Pimbblet}}]{jurek07}
{Jurek}, R.~J., {Drinkwater}, M.~J., {Francis}, P.~J., \& {Pimbblet}, K.~A.
  2007, ArXiv e-prints, 710

\bibitem[{{Labb{\'e}} {et~al.}(2003){Labb{\'e}}, {Franx}, {Rudnick},
  {Schreiber}, {Rix}, {Moorwood}, {van Dokkum}, {van der Werf},
  {R{\"o}ttgering}, {van Starkenburg}, {van de Wel}, {Kuijken}, \&
  {Daddi}}]{labbe03}
{Labb{\'e}}, I., {Franx}, M., {Rudnick}, G., {et~al.} 2003, \aj, 125, 1107

\bibitem[{{Leipski} {et~al.}(2007){Leipski}, {Haas}, {Siebenmorgen},
  {Meusinger}, {Albrecht}, {Cesarsky}, {Chini}, {Cutri}, {Drass}, {Huchra},
  {Ott}, \& {Wilkes}}]{leipski07b}
{Leipski}, C., {Haas}, M., {Siebenmorgen}, R., {et~al.} 2007, \aap, 473, 121

\bibitem[{{Lewis} {et~al.}(2002){Lewis}, {Cannon}, {Taylor}, {Glazebrook},
  {Bailey}, {Baldry}, {Barton}, {Bridges}, {Dalton}, {Farrell}, {Gray},
  {Lankshear}, {McCowage}, {Parry}, {Sharples}, {Shortridge}, {Smith},
  {Stevenson}, {Straede}, {Waller}, {Whittard}, {Wilcox}, \&
  {Willis}}]{lewis02}
{Lewis}, I.~J., {Cannon}, R.~D., {Taylor}, K., {et~al.} 2002, \mnras, 333, 279

\bibitem[{{Lonsdale} {et~al.}(2004){Lonsdale}, {Polletta}, {Surace}, {Shupe},
  {Fang}, {Xu}, {Smith}, {Siana}, {Rowan-Robinson}, {Babbedge}, {Oliver},
  {Pozzi}, {Davoodi}, {Owen}, {Padgett}, {Frayer}, {Jarrett}, {Masci},
  {O'Linger}, {Conrow}, {Farrah}, {Morrison}, {Gautier}, {Franceschini},
  {Berta}, {Perez-Fournon}, {Hatziminaoglou}, {Afonso-Luis}, {Dole}, {Stacey},
  {Serjeant}, {Pierre}, {Griffin}, \& {Puetter}}]{lonsdale04}
{Lonsdale}, C., {Polletta}, M.~d.~C., {Surace}, J., {et~al.} 2004, \apjs, 154,
  54

\bibitem[{{Lonsdale} {et~al.}(2003){Lonsdale}, {Smith}, {Rowan-Robinson},
  {Surace}, {Shupe}, {Xu}, {Oliver}, {Padgett}, {Fang}, {Conrow},
  {Franceschini}, \& et. al.}]{lonsdale03}
{Lonsdale}, C.~J., {Smith}, H.~E., {Rowan-Robinson}, M., {et~al.} 2003, \pasp,
  115, 897

\bibitem[{{Maddox} \& {Hewett}(2006)}]{maddox06}
{Maddox}, N. \& {Hewett}, P.~C. 2006, \mnras, 367, 717

\bibitem[{{Masci} {et~al.}(1998){Masci}, {Webster}, \& {Francis}}]{masci98}
{Masci}, F.~J., {Webster}, R.~L., \& {Francis}, P.~J. 1998, \mnras, 301, 975

\bibitem[{Monet {et~al.}(2003)Monet, Levine, \& Canzian}]{monet03}
Monet, D., Levine, S., \& Canzian, B. 2003, \aj, 125, 984

\bibitem[{Nakos(2007)}]{nakos07}
Nakos, T. 2007, PhD thesis, University of Li\`ege, Belgium

\bibitem[{Persson {et~al.}(2002)Persson, Murphy, Gunnels, Birk, A, \&
  Koch}]{persson02}
Persson, S., Murphy, D., Gunnels, S., {et~al.} 2002, \aj, 124, 619

\bibitem[{Persson {et~al.}(1998)Persson, Murphy, Krzeminski, Roth, \&
  Rieke}]{persson98}
Persson, S., Murphy, D., Krzeminski, W., Roth, M., \& Rieke, M. 1998, \aj, 116,
  2745

\bibitem[{Pickles(1998)}]{pickles98}
Pickles, A. 1998, \pasp, 110, 863

\bibitem[{{Pierre} {et~al.}(2001){Pierre}, {Lidman}, {Hunstead}, {Alloin},
  {Casali}, {Cesarsky}, {Chanial}, {Duc}, {Fadda}, {Flores}, {Madden}, \&
  {Vigroux}}]{pierre01}
{Pierre}, M., {Lidman}, C., {Hunstead}, R., {et~al.} 2001, \aap, 372, L45

\bibitem[{{Pierre} {et~al.}(2004){Pierre}, {Valtchanov}, {Altieri}, {Andreon},
  {Bolzonella}, {Bremer}, {Disseau}, {Dos Santos}, {Gandhi}, {Jean}, {Pacaud},
  {Read}, {Refregier}, {Willis}, {Adami}, {Alloin}, {Birkinshaw}, {Chiappetti},
  {Cohen}, {Detal}, {Duc}, {Gosset}, {Hjorth}, {Jones}, {LeFevre}, {Lonsdale},
  {Maccagni}, {Mazure}, {McBreen}, {McCracken}, {Mellier}, {Ponman},
  {Quintana}, {Rottgering}, {Smette}, {Surdej}, {Starck}, {Vigroux}, \&
  {White}}]{pierre04}
{Pierre}, M., {Valtchanov}, I., {Altieri}, B., {et~al.} 2004, Journal of
  Cosmology and Astro-Particle Physics, 9, 11

\bibitem[{{Polletta} {et~al.}(2007){Polletta}, {Tajer}, {Maraschi},
  {Trinchieri}, {Lonsdale}, {Chiappetti}, {Andreon}, {Pierre}, {Le F{\`e}vre},
  {Zamorani}, {Maccagni}, {Garcet}, {Surdej}, {Franceschini}, {Alloin},
  {Shupe}, {Surace}, {Fang}, {Rowan-Robinson}, {Smith}, \&
  {Tresse}}]{polletta07}
{Polletta}, M., {Tajer}, M., {Maraschi}, L., {et~al.} 2007, \apj, 663, 81

\bibitem[{{Richards} {et~al.}(2003){Richards}, {Hall}, {Vanden Berk},
  {Strauss}, {Schneider}, {Weinstein}, {Reichard}, {York}, {Knapp}, {Fan},
  {Ivezi{\'c}}, {Brinkmann}, {Budav{\'a}ri}, {Csabai}, \&
  {Nichol}}]{richards03}
{Richards}, G.~T., {Hall}, P.~B., {Vanden Berk}, D.~E., {et~al.} 2003, \aj,
  126, 1131

\bibitem[{{Richards} {et~al.}(2006){Richards}, {Lacy}, {Storrie-Lombardi},
  {Hall}, {Gallagher}, {Hines}, {Fan}, {Papovich}, {Vanden Berk}, {Trammell},
  {Schneider}, {Vestergaard}, {York}, {Jester}, {Anderson}, {Budav{\'a}ri}, \&
  {Szalay}}]{richards06}
{Richards}, G.~T., {Lacy}, M., {Storrie-Lombardi}, L.~J., {et~al.} 2006, \apjs,
  166, 470

\bibitem[{{Rieke} {et~al.}(2004){Rieke}, {Young}, {Engelbracht}, {Kelly},
  {Low}, \& et~al.}]{rieke04}
{Rieke}, G.~H., {Young}, E.~T., {Engelbracht}, C.~W., {et~al.} 2004, \apjs,
  154, 25

\bibitem[{{Sandage}(1965)}]{sandage65}
{Sandage}, A. 1965, \apj, 141, 1560

\bibitem[{{Schmidt} \& {Green}(1983)}]{schmidt83}
{Schmidt}, M. \& {Green}, R.~F. 1983, \apj, 269, 352

\bibitem[{{Sharp} {et~al.}(2002){Sharp}, {Sabbey}, {Vivas}, {Oemler},
  {McMahon}, {Hodgkin}, \& {Coppi}}]{sharp02}
{Sharp}, R.~G., {Sabbey}, C.~N., {Vivas}, A.~K., {et~al.} 2002, \mnras, 337,
  1153

\bibitem[{{Skrutskie} {et~al.}(2006){Skrutskie}, {Cutri}, {Stiening},
  {Weinberg}, {Schneider}, {Carpenter}, {Beichman}, {Capps}, {Chester},
  {Elias}, {Huchra}, {Liebert}, {Lonsdale}, {Monet}, {Price}, {Seitzer},
  {Jarrett}, {Kirkpatrick}, {Gizis}, {Howard}, {Evans}, {Fowler}, {Fullmer},
  {Hurt}, {Light}, {Kopan}, {Marsh}, {McCallon}, {Tam}, {Van Dyk}, \&
  {Wheelock}}]{skrutskie06}
{Skrutskie}, M.~F., {Cutri}, R.~M., {Stiening}, R., {et~al.} 2006, \aj, 131,
  1163

\bibitem[{{Smail} {et~al.}(2008){Smail}, {Sharp}, {Swinbank}, {Akiyama},
  {Ueda}, {Foucaud}, {Almaini}, \& {Croom}}]{smail08}
{Smail}, I., {Sharp}, R., {Swinbank}, A.~M., {et~al.} 2008, \mnras, 823

\bibitem[{{Smith} {et~al.}(2002){Smith}, {Tucker}, {Kent}, {Richmond},
  {Fukugita}, \& {Ichikawa}}]{smith02}
{Smith}, J.~A., {Tucker}, D.~L., {Kent}, S., {et~al.} 2002, \aj, 123, 2121

\bibitem[{{Str{\"u}der} {et~al.}(2001){Str{\"u}der}, {Briel}, {Dennerl},
  {Hartmann}, {Kendziorra}, {Meidinger}, {Pfeffermann}, {Reppin}, {Aschenbach},
  {Bornemann}, {Br{\"a}uninger}, {Burkert}, {Elender}, {Freyberg}, {Haberl},
  {Hartner}, {Heuschmann}, {Hippmann}, {Kastelic}, {Kemmer}, {Kettenring},
  {Kink}, {Krause}, {M{\"u}ller}, {Oppitz}, {Pietsch}, {Popp}, {Predehl},
  {Read}, {Stephan}, {St{\"o}tter}, {Tr{\"u}mper}, {Holl}, {Kemmer}, {Soltau},
  {St{\"o}tter}, {Weber}, {Weichert}, {von Zanthier}, {Carathanassis}, {Lutz},
  {Richter}, {Solc}, {B{\"o}ttcher}, {Kuster}, {Staubert}, {Abbey}, {Holland},
  {Turner}, {Balasini}, {Bignami}, {La Palombara}, {Villa}, {Buttler},
  {Gianini}, {Lain{\'e}}, {Lumb}, \& {Dhez}}]{struder01}
{Str{\"u}der}, L., {Briel}, U., {Dennerl}, K., {et~al.} 2001, \aap, 365, L18

\bibitem[{{Surace}(2004)}]{surace04}
{Surace}, J.~A. e.~a. 2004, in Bulletin of the American Astronomical Society,
  Vol.~36, Bulletin of the American Astronomical Society, 1450

\bibitem[{{Tajer} {et~al.}(2007){Tajer}, {Polletta}, {Chiappetti}, {Maraschi},
  {Trinchieri}, {Maccagni}, {Andreon}, {Garcet}, {Surdej}, {Pierre}, {Le
  F{\`e}vre}, {Franceschini}, {Lonsdale}, {Surace}, {Shupe}, {Fang},
  {Rowan-Robinson}, {Smith}, \& {Tresse}}]{tajer07}
{Tajer}, M., {Polletta}, M., {Chiappetti}, L., {et~al.} 2007, \aap, 467, 73

\bibitem[{{Tedds} {et~al.}(2006){Tedds}, {Page}, \& {XMM-Newton Survey Science
  Centre}}]{tedds06}
{Tedds}, J.~A., {Page}, M.~J., \& {XMM-Newton Survey Science Centre}. 2006, in
  ESA Special Publication, Vol. 604, The X-ray Universe 2005, ed. A.~{Wilson},
  843--+

\bibitem[{{Turner} {et~al.}(2001){Turner}, {Abbey}, {Arnaud}, {Balasini},
  {Barbera}, \& {Belsole}}]{turner01}
{Turner}, M.~J.~L., {Abbey}, A., {Arnaud}, M., {et~al.} 2001, \aap, 365, L27

\bibitem[{{Warren} {et~al.}(2000){Warren}, {Hewett}, \& {Foltz}}]{warren00}
{Warren}, S.~J., {Hewett}, P.~C., \& {Foltz}, C.~B. 2000, \mnras, 312, 827

\bibitem[{{Webster} {et~al.}(1995){Webster}, {Francis}, {Peterson},
  {Drinkwater}, \& {Masci}}]{webster95}
{Webster}, R.~L., {Francis}, P.~J., {Peterson}, B.~A., {Drinkwater}, M.~J., \&
  {Masci}, F.~J. 1995, \nat, 375, 469

\bibitem[{{Weedman} {et~al.}(2006){Weedman}, {Polletta}, {Lonsdale}, {Wilkes},
  {Siana}, {Houck}, {Surace}, {Shupe}, {Farrah}, \& {Smith}}]{weedman06}
{Weedman}, D., {Polletta}, M., {Lonsdale}, C.~J., {et~al.} 2006, \apj, 653, 101

\bibitem[{{Whiting} {et~al.}(2001){Whiting}, {Webster}, \&
  {Francis}}]{whiting01}
{Whiting}, M.~T., {Webster}, R.~L., \& {Francis}, P.~J. 2001, \mnras, 323, 718

\bibitem[{{Wright} \& {Otrupcek}(1990)}]{parks90}
{Wright}, A.~E. \& {Otrupcek}, R. 1990, in PKSCAT90: Radio Source Catalogue and
  Sky Atlas, Australia Telescope National Facility

\end{thebibliography}

\begin{appendix}
\section{Details concerning the photo-z calculations}
In this appendix we present in detail the parameters used for computing
the photometric redshifts of the quasars found in our survey, for which
no spectroscopic information was available. 

HyperZ was fed with 15 templates, taken from the~\cite{polletta07} library. 
They were chosen from seven different categories of active (AGN) or non-active
(ellipticals, spirals) galaxies.  Concerning the starburst galaxies,  
the templates are those of Arp\,220, M\,82 and NGC\,6090. 
The exact number of templates per category is presented in Table~\ref{photoz-templ}. 

The minimum and maximum redshifts were 0.7 and 3.0, respectively, with a
step of $\Delta z = 0.01$. 

In the photo-z computation, the possibility of dust reddening 
was incorporated via the~\cite{calzetti00} reddening law.
The minimum and maximum reddening values were 0.0 and 0.5, respectively, with a step
of 0.1.

Constraints were also set on the absolute $M_B$-band magnitude of
the sources. We used the $g'$-band filter as a reference and computed the
approximate limits in this band considering the formula by~\cite{smith02}: 

\begin{equation}
	B = g' +0.47\,(g'-r') + 0.17
\end{equation}

with $ -28.0 \le M_B \le -22.0$ and using an average $(g'-r')_{AB}$ color index of
0.65. Based on the output of HyperZ, the brightest absolute magnitude
was found to be $M_B = -25.42$ and the faintest $M_B = -22.5$, respectively.

The cosmological parameters used in this analysis are $H_0 = 70$ km\,s$^{-1}$\,Mpc$^{-1}$,
$\Omega_{\Lambda} = 0.7, \,\Omega_{M} = 0.3$.

\begin{table}[b]
\caption[]{Distribution of the template categories used for the photometric 
redshift calculations.}
\label{photoz-templ}
\begin{tabular}{lc}\\ \hline \hline 
\rule[-3mm]{0mm}{8mm}
  Template type & Nr     \\ \hline
\rule[0mm]{-0.5mm}{4mm} 
\hspace{-1mm}
   Type-1 quasar   &  3  \\ 
   Type-2 quasar   &  2  \\ 
   Seyfert      &  2  \\ 
   Starbursts   &  3  \\      
   Ellipt.      &  3  \\
   Spirals      &  2  \\ \hline
\end{tabular}
\end{table}

\end{appendix}

\begin{landscape}
\begin{table*}
\caption[]{Sources in the LCO field with a 2dF spectroscopic identification.
Column 1 gives the identifier of the source in the spectroscopic data base.
Columns 2 and 3 give the right  ascension and declination of each source, for equinox 2000. 
Columns 4, 5 provide the spectroscopic classification of the source (Star, Quasar or
Galaxy) and its measured redshift (for the latter  two object-types), respectively. 
The typical uncertainty on the estimated redshift is on the order of 0.001 and 0.005 
for the galaxies and quasars, respectively. 
Columns $6-7$, $8-9$ and $10-11$ give the 3$''$ photometry and its error for the 
$R,z'$ and \ks-bands, respectively. 
Columns $11-12$ provide the (\rmz) and (\zmk) color index for each source, respectively. 
The five quasars not included in the matched \rzk catalog are presented at the bottom of
Table~\ref{2dF-tbl} and are marked as ``Q*".}	
\label{2dF-tbl}							      
\centering
\footnotesize
\begin{tabular}{rccccccccccrr}\hline 
\rule[-3mm]{0mm}{8mm}
 ID  & RA (J2000)                       & Dec (J2000)                       & Type   & Rdsft    &   $R$	    & $\sigma_R$  &   $z'$     & $\sigma_{z'}$ &   \ks	  & $\sigma_{K_s}$  &	\rmz	& \zmk   \\ 
     & \hspace{-3mm}h m\hspace{3mm} s   &\hspace{-3mm}$^{\circ} \hspace{3mm}' \hspace{3mm}''$   &           &             &  	       &	       &	  & 	            & 	        &		  &		&	   \\ \hline \hline
\rule[0mm]{0mm}{4mm} 
\hspace{-1mm}
 528 &  2 22 55.08                   	& $-$4 12 10.58 		    &	S    &   $-$	&  19.012   &  0.002	  &  17.359    &  0.001        &  14.949   &	0.021	     &       1.654  &   2.410  \\ \hline
 735 &  2 22 44.40		     	& $-$4 33 47.02 		    &	Q    &  0.760	&  18.307   &  0.001	  &  18.510    &  0.003        &  16.309   &	0.034	     &    $-$0.203  &   2.201  \\ 
 736 &  2 22 47.90		     	& $-$4 33 30.20 		    &	Q    &  1.629	&  20.151   &  0.006	  &  20.122    &  0.013        &  17.900   &	0.128	     &       0.029  &   2.222  \\ 
  97 &  2 23 26.47		     	& $-$4 57 06.30 		    &	Q    &  0.826	&  20.994   &  0.013	  &  20.968    &  0.029        &  17.843   &	0.135	     &       0.026  &   3.125  \\ 
 100 &  2 23 54.82		     	& $-$4 48 15.19 		    &	Q    &  2.463	&  18.163   &  0.001	  &  18.371    &  0.003        &  16.092   &	0.023	     &    $-$0.208  &   2.279  \\ 
 251 &  2 24 29.14		     	& $-$4 58 08.11 		    &	Q    &  1.497	&  19.626   &  0.004	  &  19.959    &  0.011        &  17.907   &	0.101	     &    $-$0.333  &   2.052  \\ 
  23 &  2 25 14.40		     	& $-$4 47 00.38 		    &	Q    &  1.924	&  19.225   &  0.003	  &  18.456    &  0.004        &  16.918   &	0.045	     &       0.769  &   1.538  \\ 
  41 &  2 25 40.61		     	& $-$4 38 25.30 		    &	Q    &  2.483	&  19.860   &  0.005	  &  19.898    &  0.013        &  17.625   &	0.096	     &    $-$0.038  &   2.273  \\ 
  11 &  2 25 56.83		     	& $-$4 58 53.29 		    &	Q    &  1.183	&  20.947   &  0.014	  &  21.147    &  0.039        &  17.968   &	0.127	     &    $-$0.200  &   3.179  \\
  14 &  2 25 57.62		     	& $-$4 50 05.50 		    &	Q    &  2.263	&  19.454   &  0.004	  &  19.722    &  0.011        &  17.826   &	0.084	     &    $-$0.268  &   1.896  \\ \hline
 531 &  2 22 57.98	             	& $-$4 18 40.50 		    &	G    &  0.237	&  19.047   &  0.003	  &  18.112    &  0.002        &  14.674   &	0.010	     &       0.935  &   3.438  \\ 
 544 &  2 23 02.04	             	& $-$4 32 04.81 		    &	G    &  0.616	&  20.437   &  0.008	  &  20.004    &  0.012        &  16.767   &	0.040	     &       0.433  &   3.237  \\ 
 742 &  2 23 15.36	             	& $-$4 25 58.69 		    &	G    &  0.190	&  19.695   &  0.004	  &  19.423    &  0.007        &  17.099   &	0.064	     &       0.272  &   2.324  \\ 
 743 &  2 23 19.66	             	& $-$4 47 30.80 		    &	G    &  0.293	&  19.284   &  0.003	  &  18.798    &  0.004        &  15.651   &	0.045	     &       0.486  &   3.147  \\ 
  96 &  2 23 44.26	             	& $-$4 57 25.42 		    &	G    &  0.157	&  19.803   &  0.005	  &  19.576    &  0.008        &  16.635   &	0.034	     &       0.227  &   2.941  \\ 
  65 &  2 23 51.29	             	& $-$4 20 53.41 		    &	G    &  0.181	&  19.635   &  0.004	  &  19.666    &  0.009        &  17.144   &	0.056	     &    $-$0.031  &   2.522  \\ 
 125 &  2 24 18.79	             	& $-$5 01 20.89 		    &	G    &  0.458	&  20.769   &  0.011	  &  20.589    &  0.020        &  18.177   &	0.141	     &       0.180  &   2.412  \\ 
 248 &  2 25 21.07	             	& $-$4 39 49.90 		    &	G    &  0.265	&  19.416   &  0.003	  &  18.941    &  0.005        &  16.074   &	0.048	     &       0.475  &   2.867  \\ 
 243 &  2 25 24.70	             	& $-$4 40 44.69 		    &	G    &  0.263	&  19.044   &  0.003	  &  18.448    &  0.004        &  15.412   &	0.017	     &       0.596  &   3.036  \\ 
  46 &  2 25 31.39	             	& $-$4 42 20.30 		    &	G    &  0.209	&  20.015   &  0.006	  &  19.929    &  0.013        &  17.806   &	0.074	     &       0.086  &   2.123  \\ 
  42 &  2 25 32.02	             	& $-$4 43 46.20 		    &	G    &  0.314	&  20.224   &  0.007	  &  19.731    &  0.011        &  16.929   &	0.048	     &       0.493  &   2.802  \\ 
 357 &  2 25 58.87	             	& $-$5 00 54.50 		    &	G    &  0.148	&  18.324   &  0.001	  &  17.921    &  0.002        &  15.106   &	0.029	     &       0.403  &   2.815  \\ \hline
 530 &  2 22 49.63                   	& $-$4 13 52.97 		    &	Q*   &  1.566	&  20.301   &  0.007	  &  20.461    &  0.018        &  17.793   &	0.087	     &    $-$0.160  &   2.668 \\       
 281 &  2 23 51.10                   	& $-$4 47 29.76 		    &	Q*   &  2.164	&  20.065   &  0.005	  &  20.311    &  0.015        &  18.001   &	0.101	     &    $-$0.246  &   2.310 \\
  91 &  2 23 58.66                   	& $-$4 53 51.40 		    &	Q*   &  2.275	&  19.866   &  0.004	  &  20.084    &  0.012        &  17.509   &	0.066	     &    $-$0.218  &   2.575 \\
 271 &  2 24 13.46                   	& $-$4 52 10.27 		    &	Q*   &  2.487	&  20.646   &  0.009	  &  20.913    &  0.026        &  18.097   &	0.110	     &    $-$0.267  &   2.816 \\
 375 &  2 25 37.03                   	& $-$5 01 09.41 		    &	Q*   &  1.937	&  19.968   &  0.005	  &  19.953    &  0.013        &  18.076   &	0.106	     &       0.015  &   1.877 \\ \hline
\end{tabular}
\end{table*}													      
\end{landscape}														      

\begin{landscape}
\begin{table*}
\caption[]{Astrometry and optical photometry for the 19 type-1 and type-2 newly discovered quasars
in a sub-region of the XMM-LSS field.
Column 1 gives the identifier of the source.  
Columns 2 and 3 provide the right  ascension and declination for the equinox 2000, respectively. 
Column 4 provides the photometric redshift of the source.
Columns $5-6$, $7-8$, $9-10$ and $11-12$ give the 3$''$  photometry (Vega) and 
its error for the CFHTLS $u^*, g', r', i'$ and $z'$-bands, respectively.
\\
{\sl Notes on individual objects:}
Objects 1S and 2S are the quasars found among the point-like sources, using the SWIRE
color-diagram. A third object found in the same way (see text for more
details) coincides with the 2dF quasar \#23 (see Table~\ref{2dF-tbl}).
Object $\#$33 is the only QSO with typical SED features of a type-2 object. Due to the lack of 
spectroscopic information, the objects presented in this table were characterized as type-1 quasars
based on their SED. For objects 2,3, 20 and 27 we have recently obtained VIMOS spectra, which have
indeed confirmed the SED classification. The measured spectroscopic redshift is 1.165, 1.900, 1.085 and 2.148 respectively.
}	
\label{type1-type2-qso-p1}							      
\centering
\footnotesize
\begin{tabular}{rccccccccccccc}\hline 
\rule[-3mm]{0mm}{8mm}
   ID    &  RA (J2000)                          &  Dec (J2000)                                         	  &  Redshift   &     $u^*$  & $\sigma_{u^*}$ &     $g'$     & $\sigma_{g'}$  &	   $r'$    & $\sigma_{r'}$  &    $i'$     & $\sigma_{i'}$  &   $z'$     &  $\sigma_{z'}$   \\  
         &  \hspace{-3mm}h m\hspace{3mm} s      & \hspace{-3mm}$^{\circ} \hspace{3mm}' \hspace{3mm}''$ 	  &             &     (Vega) &	              &   (Vega)     &		      &	   (Vega)  &		    &    (Vega)   &		   &  (Vega)    &		   \\ \hline \hline
\rule[0mm]{0mm}{4mm} 
\hspace{-1mm}
    2	 & 	2  22  42.54    		 &  $-$4  30  18.49     				  &    1.110	&    21.371   &     0.020     &     21.304   &     0.020      &    20.540   &	  0.013     &	 20.527   &	0.014	   &	20.326   &     0.018   \\
    3	 & 	2  22  42.92    		 &  $-$4  33  14.64     				  &    1.513	&    20.511   &     0.013     &     20.995   &     0.013      &    20.445   &	  0.013     &	 20.113   &	0.011	   &	19.822   &     0.012   \\
   17	 & 	2  22  58.89    		 &  $-$4  58  52.41     				  &    0.710	&    19.693   &     0.009     &     20.006   &     0.009      &    19.217   &	  0.007     &	 18.790   &	0.006	   &	18.376   &     0.005   \\
   19	 & 	2  23  04.15    		 &  $-$4  44  35.40     				  &    2.262	&    20.355   &     0.011     &     20.656   &     0.011      &    20.138   &	  0.010     &	 19.710   &	0.009	   &	19.323   &     0.009   \\
   20	 & 	2  23  06.05    		 &  $-$4  33  23.93     				  &    0.910	&    20.381   &     0.011     &     20.693   &     0.011      &    20.292   &	  0.012     &	 20.156   &	0.011	   &	19.779   &     0.012   \\
   27	 & 	2  23  25.63    		 &  $-$4  22  54.00     				  &    1.574	&    22.246   &     0.032     &     22.426   &     0.032      &    21.360   &	  0.021     &	 21.147   &	0.021	   &	20.756   &     0.026   \\
   33	 & 	2  23  36.87    		 &  $-$4  30  51.84     				  &    1.679	&    23.928   &     0.093     &     23.817   &     0.093      &    22.509   &	  0.044     &	 21.181   &	0.021	   &	20.479   &     0.020   \\
   37	 & 	2  23  40.78    		 &  $-$4  22  55.38     				  &    0.910	&    19.666   &     0.009     &     19.818   &     0.009      &    19.188   &	  0.007     &	 19.085   &	0.007	   &	18.717   &     0.006   \\
   45	 & 	2  23  50.77    		 &  $-$4  31  58.25     				  &    1.158	&    19.731   &     0.010     &     19.922   &     0.010      &    19.200   &	  0.006     &	 18.810   &	0.006	   &	18.660   &     0.006   \\
   2S	 & 	2  23  52.18    		 &  $-$4  30  31.81     				  &    2.113	&    19.679   &     0.012     &     19.975   &     0.012      &    19.514   &	  0.007     &	 19.149   &	0.007	   &	18.859   &     0.007   \\
   47	 & 	2  23  52.54    		 &  $-$4  18  21.65     				  &    2.361	&    20.712   &     0.014     &     20.750   &     0.014      &    20.125   &	  0.010     &	 19.747   &	0.010	   &	19.309   &     0.009   \\
   57	 & 	2  24  16.65    		 &  $-$4  56  43.02     				  &    1.481	&    21.270   &     0.019     &     21.381   &     0.019      &    20.663   &	  0.014     &	 20.373   &	0.013	   &	19.965   &     0.014   \\
   1S	 & 	2  24  24.17    		 &  $-$4  32  29.85     				  &    1.901	&    18.842   &     0.006     &     19.207   &     0.006      &    18.841   &	  0.005     &	 18.449   &	0.005	   &	18.318   &     0.005   \\
   78	 & 	2  25  15.34    		 &  $-$4  40  08.86     				  &    2.295	&    19.712   &     0.009     &     19.966   &     0.009      &    19.554   &	  0.009     &	 19.140   &	0.007	   &	19.017   &     0.008   \\
   80	 & 	2  25  34.82    		 &  $-$4  24  01.69     				  &    0.910	&    20.595   &     0.014     &     20.985   &     0.014      &    20.524   &	  0.012     &	 20.233   &	0.012	   &	19.767   &     0.013   \\ 
   81	 & 	2  25  37.16    		 &  $-$4  21  32.85     				  &    0.974	&    19.541   &     0.009     &     19.694   &     0.009      &    19.130   &	  0.006     &	 18.981   &	0.006	   &	18.740   &     0.007   \\
   83	 & 	2  25  37.55    		 &  $-$4  54  40.29     				  &    2.445	&    21.448   &     0.021     &     21.111   &     0.021      &    20.115   &	  0.017     &	 19.586   &	0.010	   &	19.324   &     0.011   \\
   84	 & 	2  25  39.36    		 &  $-$4  22  28.02     				  &    1.007	&    21.025   &     0.017     &     21.390   &     0.017      &    20.816   &	  0.014     &	 20.486   &	0.014	   &	19.948   &     0.014   \\
   85	 & 	2  25  55.43    		 &  $-$4  39  18.16     				  &    0.948	&    20.769   &     0.015     &     21.027   &     0.015      &    20.371   &	  0.011     &	 20.166   &	0.012	   &	19.806   &     0.012   \\  \hline
\end{tabular}																									   
\end{table*}													      												    
\end{landscape}														      											   

\begin{landscape}
\begin{table*}
\caption[]{NIR and MIR properties of the type-1 and type-2 quasars presented in Table~\ref{type1-type2-qso-p1}.
Column 1 gives the source identification. 
Columns $2-3$, $4-5$ and $6-7$  give the 3$''$ photometry and its error for the 
$R,z'$ (CTIO) and \ks-bands, respectively (Vega magnitudes). 
It should be noted that althought the cut-off magnitude limit of the \ks-band catalog is
$K_s = 18$, some of the \ks-band magnitudes might appear fainter.
This results from the fact that the cut-off limit has been implemented using 
SExtractor's Kron photometry, while the magnitudes presented here are computed 
based on the 3$''$ photometry.
Columns $8-9$ $10-11, 12-13$ and $14-15$  provide the photometry 
and the corresponding error for the four IRAC bands (3.6\micron, 4.5\micron, 5.8\micron\, and 8.0\micron),
in $\mu$Jy units.
Columns $16-17$ provide the photometry and its photometric error for the MIPS\,24\micron\ photometric band.
}	
\label{type1-agn-qso-p2}							      
\centering
\scriptsize
\begin{tabular}{rrrrrrrrrrrccrrrr}\hline 
\rule[-3mm]{0mm}{8mm}
 ID     &         $R$    &  $\sigma_R$  & $z'_{ctio}$  & $\sigma_{z'}$  &   \ks    &  $\sigma_{K_s}$ &  Flux$_{3.6}$  &  $\sigma_{F(3.6)}$   & Flux$_{4.5}$  & $\sigma_{F(4.5)}$ & Flux$_{5.8}$   & $\sigma_{F(5.8)}$  & Flux$_{8.0}$  & $\sigma_{F(8.0)}$  & Flux$_{24}$   & $\sigma_{F(24)}$   \\  
        &     (Vega)     &	        &    (Vega)     &		&  (Vega)  &		     &($\mu$Jy)       &			     &	($\mu$Jy)    &		         &  ($\mu$Jy)     &		       &  ($\mu$Jy)    &  		    &  ($\mu$Jy)    &		         \\ \hline \hline
\rule[0mm]{0mm}{4mm} 
\hspace{-1mm}
     2  &   	20.677   &    0.010	&  20.484      &     0.030	&  18.065  &	     0.195   &     67.19      & 	1.03	     &     85.41     &       1.20	 &    103.62	  &	   5.70        &    129.28     &	5.74	    &	  403.71    &	    25.88    	 \\
     3  &   	20.610   &    0.010	&  20.080      &     0.021	&  18.316  &	     0.195   &     63.27      & 	1.24	     &    101.75     &       1.79	 &    177.03	  &	   7.37        &    252.46     &	8.17	    &	  673.71    &	    26.94    	 \\
    17  &   	19.337   &    0.003	&  18.774      &     0.006	&  16.407  &	     0.029   &    291.91      & 	2.32	     &    318.66     &       2.64	 &    396.30	  &	   8.54        &    591.00     &	8.88	    &	 2844.53    &	    24.61    	 \\
    19  &   	20.446   &    0.008	&  19.811      &     0.017	&  18.037  &	     0.109   &     60.36      & 	0.93	     &     74.04     &       1.17	 &    113.14	  &	   5.01        &    178.77     &	5.66	    &	  566.74    &	    24.26    	 \\
    20  &   	20.394   &    0.008	&  20.084      &     0.021	&  17.394  &	     0.073   &    182.39      & 	1.52	     &    219.32     &       2.29	 &    275.59	  &	   6.95        &    317.70     &	8.28	    &	  915.17    &	    24.54    	 \\
    27  &   	21.415   &    0.020	&  20.943      &     0.046	&  18.171  &	     0.217   &     66.69      & 	1.26	     &     73.79     &       1.38	 &     55.10	  &	   6.75        &    103.54     &	6.42	    &	    0.00    &	     0.00    	 \\
    33  &   	22.305   &    0.046	&  20.630      &     0.036	&  18.035  &	     0.150   &     81.77      & 	1.36	     &     87.79     &       1.72	 &    139.38	  &	   7.21        &    237.20     &	8.14	    &	 1329.34    &	    26.89    	 \\
    37  &   	19.473   &    0.004	&  19.049      &     0.008	&  17.351  &	     0.065   &    270.60      & 	1.81	     &    417.24     &       2.61	 &    611.26	  &	   7.73        &    915.69     &	8.34	    &	 3005.32    &	    24.35    	 \\
    45  &   	19.394   &    0.003	&  18.908      &     0.007	&  17.223  &	     0.057   &    177.91      & 	1.83	     &    254.97     &       2.15	 &    351.71	  &	   8.31        &    471.64     &	7.66	    &	  502.97    &	    27.92    	 \\
    2S  &   	19.847   &    0.005	&  19.170      &     0.009	&  17.820  &	     0.099   &     96.81      & 	1.19	     &    130.62     &       1.70	 &    239.63	  &	   6.41        &    405.71     &	7.52	    &	 1805.34    &	    24.71    	 \\   
    47  &   	20.267   &    0.007	&  19.582      &     0.013	&  17.739  &	     0.090   &     79.32      & 	1.09	     &    103.61     &       1.45	 &    194.90	  &	   6.93        &    278.18     &	6.67	    &	  689.74    &	    27.95    	 \\
    57  &   	20.696   &    0.010	&  20.259      &     0.024	&  17.539  &	     0.082   &    143.05      & 	1.72	     &    161.52     &       2.04	 &    173.15	  &	   7.40        &    253.27     &	8.19	    &	  793.37    &	    23.52    	 \\
    1S  &   	18.914   &    0.002	&  18.382      &     0.005	&  17.400  &	     0.146   &    147.40      & 	1.52	     &    219.71     &       2.26	 &    378.36	  &	   7.73        &    546.59     &	8.70	    &	 1086.67    &	    23.37    	 \\   
    78  &   	19.839   &    0.005	&  19.345      &     0.013	&  18.115  &	     0.212   &     86.71      & 	1.36	     &    138.69     &       2.14	 &    248.54	  &	   7.88        &    440.74     &	8.65	    &	  618.96    &	    25.88    	 \\
    80  &   	20.965   &    0.014	&  20.394      &     0.029	&  17.882  &	     0.100   &    120.90      & 	1.55	     &    116.97     &       1.84	 &    155.51	  &	   7.34        &    191.28     &	8.28	    &	  701.69    &	    29.69    	 \\
    81  &   	19.196   &    0.003	&  19.012      &     0.008	&  17.546  &	     0.138   &    172.90      & 	1.79	     &    224.93     &       2.34	 &    365.75	  &	   8.42        &    556.36     &	8.85	    &	 1113.69    &	    28.03    	 \\
    83  &   	20.399   &    0.009	&  19.717      &     0.018	&  17.658  &	     0.089   &     43.84      & 	0.96	     &     43.20     &       1.49	 &	0.00	  &	   0.00        &    178.43     &	8.20	    &	    0.00    &	     0.00    	 \\
    84  &   	20.889   &    0.013	&  20.338      &     0.028	&  17.756  &	     0.136   &    105.96      & 	1.48	     &     92.05     &       1.74	 &     94.04	  &	   7.19        &      0.00     &	0.00	    &	    0.00    &	     0.00    	 \\
    85  &   	20.660   &    0.010	&  20.221      &     0.028	&  18.086  &	     0.158   &    134.12      & 	1.62	     &    173.74     &       2.08	 &    241.01	  &	   7.76        &    297.63     &	8.39	    &	  752.02    &	    26.36    	 \\ \hline
\end{tabular}
\end{table*}													      
\end{landscape}														      

\end{document}